\documentclass[conference]{IEEEtran}
\IEEEoverridecommandlockouts
\usepackage{cite}
\usepackage{amsmath,amssymb,amsfonts}
\usepackage{graphicx}
\usepackage{textcomp}
\def\BibTeX{{\rm B\kern-.05em{\sc i\kern-.025em b}\kern-.08em
    T\kern-.1667em\lower.7ex\hbox{E}\kern-.125emX}}

\usepackage{algorithm}
\usepackage{algpseudocode}
\usepackage{xcolor}
\usepackage{caption,subcaption}
\usepackage{booktabs}
\usepackage{color, colortbl}
\definecolor{Gray}{gray}{0.93}
\usepackage{amssymb}
\usepackage{pifont}
\newcommand{\cmark}{\ding{51}}%
\newcommand{\xmark}{\ding{55}}%
\newcommand{\atom}{\textsc{Atom} } 
\newcommand{\atomnospace}{\textsc{Atom} }
\algnewcommand{\algorithmicand}{\textbf{ and }}
\algnewcommand{\algorithmicor}{\textbf{ or }}
\algnewcommand{\OR}{\algorithmicor}
\algnewcommand{\AND}{\algorithmicand}
\algnewcommand{\var}{\texttt}

\usepackage{url}

\usepackage{hyperref}
\hypersetup{colorlinks,allcolors=black}
\usepackage{graphicx}

\begin{document}
\title{\atomnospace: Asynchronous Training of Massive Models for Deep Learning in a Decentralized Environment}

\author{
    \IEEEauthorblockN{Xiaofeng Wu}
    \IEEEauthorblockA{
        \textit{City University of Macau}\\
        Macau, China\\
        xiaofengwu@cityu.edu.mo
    }
    \and
    \IEEEauthorblockN{Jia Rao}
    \IEEEauthorblockA{
        \textit{The University of Texas at Arlington}\\
        Arlington, USA\\
        jia.rao@uta.edu
    }
    \and
    \IEEEauthorblockN{Wei Chen}
    \IEEEauthorblockA{
        \textit{Nvidia Corporation}\\
        Location of Nvidia Corporation\\
        weich@nvidia.com
    }
}

\maketitle

\thispagestyle{plain}
\pagestyle{plain}

\begin{abstract}
The advent of the Transformer architecture has propelled the growth of natural language processing (NLP) models, leading to remarkable achievements in numerous NLP tasks. Yet, the absence of specialized hardware like expansive GPU memory and high-speed interconnects poses challenges for training large-scale models. This makes it daunting for many users to experiment with pre-training and fine-tuning large language models (LLMs). In this study, we introduce \atom, a resilient distributed training framework designed for asynchronous training of vast models in a decentralized setting using cost-effective hardware, including consumer-grade GPUs and Ethernet. Unlike conventional model partitioning methods that distribute sub-models across GPUs, \atom aims to accommodate a complete LLM on one host (peer) through seamlessly model swapping and concurrently trains multiple copies across various peers to optimize training throughput. Through static analysis, \atom identifies the best model partitioning strategy and flawlessly merges model execution with swapping. Key benefits of \atom include: Avoiding the central point of failure found in pipeline parallelism methods. Demonstrating superior performance and scalability compared to closely-integrated pipeline parallelism in slower networks. Our experiments using different GPT-3 model configurations reveal that, in scenarios with suboptimal network connections, \atom can enhance training efficiency up to $20 \times$ when juxtaposed with the state-of-the-art decentralized pipeline parallelism approaches.
\end{abstract}

\begin{IEEEkeywords}
Large Language Models (LLMs),
Decentralized Training,
Deep Learning
\end{IEEEkeywords}

\section{Introduction}
Deep Neural Networks have advanced significantly in recent years with the use of deeper layers and larger model capacity, leading to improved performance in a variety of machine learning tasks such as computer vision~\cite{he2016deep}, natural language understanding~\cite{brown2020language}, structural biology, and drug discovery~\cite{jumper2021highly}. The introduction of Transformer models~\cite{vaswani2017attention} has further enhanced the capabilities of DNNs, enabling the development of large language models (LLMs) that can be applied to a wide range of applications, such as the generation of programming code with OpenAI Codex~\cite{chen2021evaluating} and GitHub Copilot~\cite{copilot}.

Recent advances in natural language processing (NLP) have led to the development of large models such as BERT-Large~\cite{devlin2018bert}, GPT-2~\cite{radford2019language}, Megatron-LM~\cite{shoeybi2019megatron}, T5~\cite{raffel2019exploring}, and GPT-3~\cite{brown2020language}, with model sizes ranging from 0.3 billion to 175 billion parameters. These models require significant resources to train from scratch, such as the use of thousands of accelerators and large text datasets. For example, GPT-3 175B, which has 175 billion parameters, was trained using 45 TB of text data and required over one month to complete training. This is a significant increase in model size compared to image recognition models such as ResNet-50~\cite{he2016deep} which has 25.6 million parameters, outpacing the development of computational, networking, and storage hardware.

While training large language models from scratch can be a challenging task for individual users and small- or medium-sized institutions and companies, there is an increasing effort to make pre-trained LLMs publicly available~\cite{gpt-3-openai, gpt-3-meta, zhang2022opt}. This allows users to experiment with pre-trained models and adapt them to their own datasets. However, LLMs are too large to fit on a single GPU, so researchers have been exploring ways to split the model and perform training in a distributed manner. This includes techniques such as model parallelism~\cite{shazeer2018mesh, lepikhin2020gshard, xu2021gspmd, athlur2022varuna} and pipeline parallelism~\cite{huang2019gpipe, narayanan2019pipedream, eliad2021fine} which divide the model into sub-models and train them on multiple GPUs. These approaches enable the use of LLMs by leveraging the device memory on distributed GPUs.

\begin{figure}[ptb!]
\setlength{\belowcaptionskip}{-0.4cm}
\centering
\includegraphics[width=1.\columnwidth]{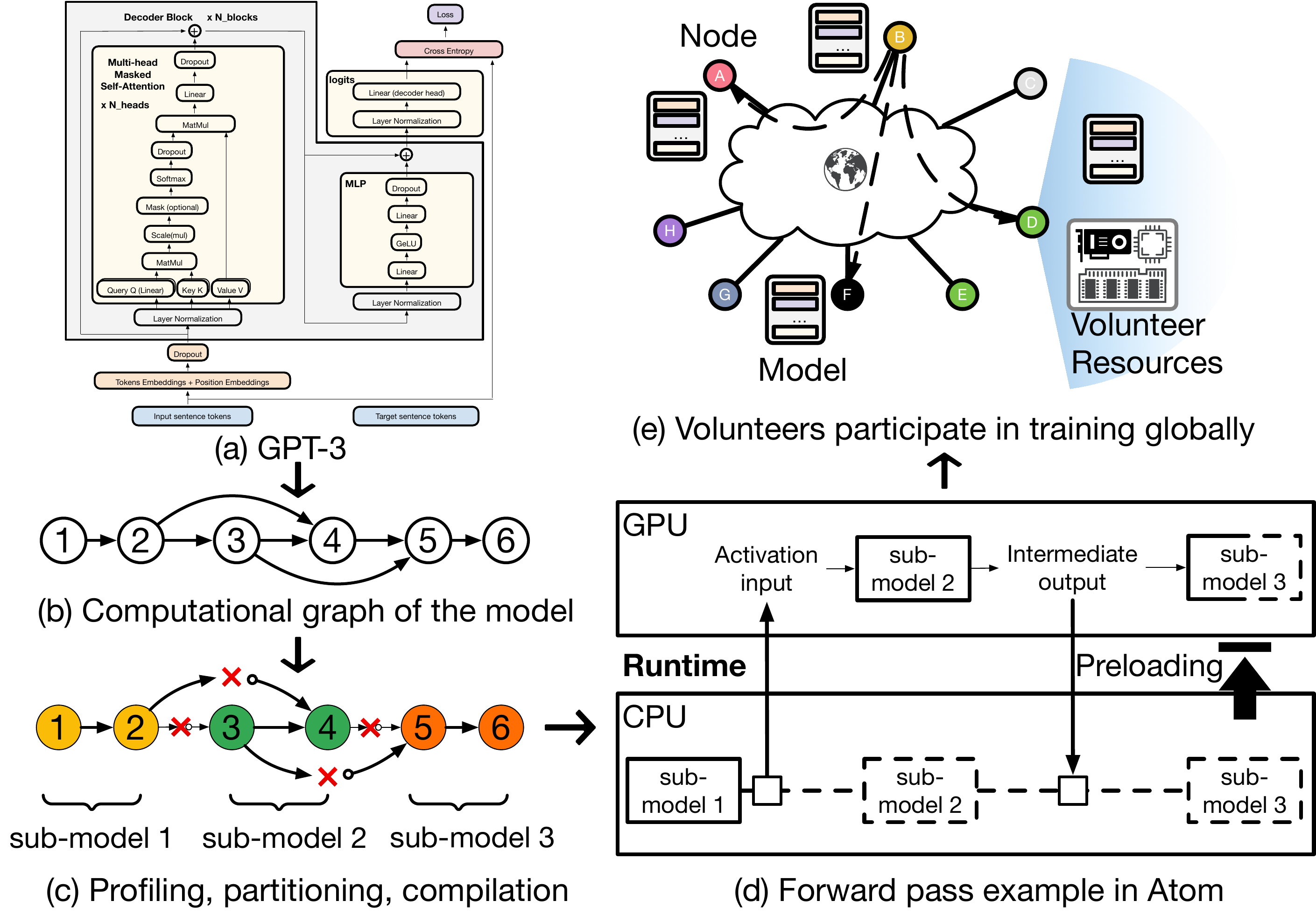}
\caption{The figure illustrates the process and functionality of the Atom system for training large language models. The system is composed of several components: (a) the user model, (b) the automated partitioning of the computation graph, (c) the compilation of sub-models code, (d) the Atom runtime for swapping sub-models between host CPU and GPU, and (e) the ability for independent participants to join and leave the training process without disrupting it. The overall goal of the system is to efficiently and effectively train large language models while addressing the challenges of limited resources and flexibility.}
\label{figs:atom_design_introduction}
\end{figure}

\begin{table*}[t]
\centering
\caption{Comparison between different large model training schemes.}
\begin{tabular}{cccccc}
\toprule[1.5pt]
\multicolumn{1}{c}{Training Schemes} & 
\multicolumn{1}{c}{Support LLMs} & 
\multicolumn{1}{c}{Host memory extension} & 
\multicolumn{1}{c}{Slow network prone} & 
\multicolumn{1}{c}{Elastic} & 
\multicolumn{1}{c}{Fault tolerance} \\
\midrule
\rowcolor{Gray}
\begin{tabular}{@{}c@{}}
Pipeline Parallelism approaches: \\ 
{GPipe}~\cite{huang2019gpipe}, 
{PipeDream}~\cite{narayanan2019pipedream}, 
FTPipe~\cite{eliad2021fine}, \\ 
Megatron-LM~\cite{shoeybi2019megatron}, 
Terapipe~\cite{li2021terapipe}  
\end{tabular}
& \cmark    & \xmark    & \xmark    & \xmark    & \xmark \\
\begin{tabular}{@{}c@{}}
DeepSpeed~\cite{deepspeed}, 
ZeRO-Offload~\cite{ZeRO-Offload},\\
FSDP~\cite{fsdp},
SwapAdvisor~\cite{huang2020swapadvisor},\\
HUVM~\cite{choi2022memory},
vDNN~\cite{rhu2016vdnn},
Capuchin~\cite{peng2020capuchin}
\end{tabular}
& \cmark    & \cmark    & \xmark    & \xmark    & \xmark \\
\rowcolor{Gray}
\begin{tabular}{@{}c@{}}
Alpa~\cite{zheng2022alpa}, 
GSPMD~\cite{xu2021gspmd}, \\
GShard~\cite{lepikhin2020gshard},
TF-Mesh~\cite{shazeer2018mesh}          
\end{tabular}
& \cmark    & \xmark    & \xmark    & \xmark    & \xmark \\
Ray~\cite{moritz2018ray}                
& \cmark    & \xmark    & \xmark    & \cmark    & \cmark \\ 
\rowcolor{Gray}
Hivemind~\cite{ryabinin2020towards}, 
DeDLOC~\cite{diskin2021distributed}     
& \xmark    & \xmark    & \cmark    & \cmark    & \cmark \\
Swarm~\cite{ryabinin2021swarm},
Bloom~\cite{scao2022bloom}
& \cmark    & \xmark    & \cmark    & \xmark    & \cmark \\
\rowcolor{Gray}
\textbf{Atom} (Our work)                
& \cmark    & \cmark    & \cmark    & \cmark    & \cmark \\
\bottomrule[1.5pt]
\end{tabular}
\label{table:training-schemes}
\end{table*}

Despite the efforts to make pre-trained LLMs publicly available and the development of techniques to split and train them in a distributed manner, there are still challenges for average users to adopt these approaches. First, access to a large number of GPUs to accommodate LLMs is still limited to a few large institutions. Second, tight coupling between sub-model training on different GPUs is a major issue, requiring not only high-speed interconnects for transmitting model output and intermediate data, but also creating a single point of failure. Third, in cases where nodes can dynamically join and leave, a model needs to be re-partitioned before training can be resumed, and this can lead to a restart of training. Fourth, if the network bandwidth is constrained, it can limit the training performance.

In this paper, we propose a solution for training large language models (LLMs) on a small number of commodity GPUs in a decentralized network environment with low-bandwidth and unreliable connections. Our key insight is that, although LLMs are extremely large in size, individual operators/layers, the building blocks of a computation graph representing the model, can be easily fit in a single commodity GPU. This opens up opportunities to execute the computation graph layer by layer through device-to-host memory swapping. Theoretically, even the GPT-3 175B model can fit in a single server given sufficient host memory. We explore an alternative approach to distributed training using loosely coupled GPUs, each training a complete LLM through memory swapping, to process massive mini-batches in parallel.

To this end, we present \atom, a distributed training approach that supports asynchronous training of massive-scale models in a decentralized environment. As shown in Figure~\ref{figs:atom_design_introduction}, \atom takes a computation graph of an LLM\footnote{In this paper, we focus on using GPT as the large language model (LLM) due to its widespread usage and popularity in the field of natural language processing.} as input and partitions the graph into sub-models based on a layer-by-layer profiling. \atom automatically generates an optimal schedule of the sub-models to streamline model execution and loading. \atom addresses the following challenges. {\em First}, memory swapping has been long criticized for low performance and high overhead. \atom avoids swapping overhead by asynchronously scheduling swapping and prefetching upcoming layers. To avoid GPU idleness, we also overlap the execution time of sub-models and the loading time of an upcoming sub-model. {\em Second,} searching an optimal model partitioning is difficult and time-consuming. \atom addresses this issue via a heuristic exhaustive search algorithm that integrates domain knowledge about the GPT-3 model. For elasticity and fault tolerance, \atom uses a DHT (distributed hash table) and a global batch size to coordinate the training on independent participants, which allows nodes to join and leave. 

We implement \atom using Pytorch and Hivemind~\cite{hivemind}. Evaluation results using GPT-3 model configurations and various network conditions show that \atom consistently outperforms Petals~\cite{borzunov2022petals} by using the schedule policy of GPipe~\cite{huang2019gpipe} and PipeDream~\cite{narayanan2019pipedream} by a large margin. Experiments with three types of commodity GPUs demonstrate that the loosely coupled distributed architecture of \atom is more performant and scalable than the tightly-coupled pipeline architecture, especially when the interconnect is slow.

\section{Background and Related Work}

\autoref{table:training-schemes} provides a comparison of \atom with various other distributed training schemes, emphasizing their primary advantages and disadvantages in decentralized LLM training.

\subsection{Volunteer Computing} 

Volunteer computing (VC)~\cite{ryabinin2020crowdsourced, diskin2021distributed, ryabinin2021swarm} is a distributed training paradigm to harvest geographically idle computing power and storage resources without advanced computing and networking devices. The academic institutes~\cite{NationalResearchCloud, anderson2004boinc} and communities~\cite{EleutherAI, BigScience} seek to democratize the access to large language models via collaborative training on voluntary resources. Hivemind~\cite{ryabinin2020towards, borzunov2022training, diskin2021distributed} trains a Mixture of Experts (MoE)~\cite{shazeer2017outrageously, lepikhin2020gshard} model in an asynchronous manner by volunteer device over a decentralized network. Distributed Deep Learning in Open Collaborations (DeDLOC)~\cite{diskin2021distributed} demonstrates the feasibility of collaborative training using model ALBERT~\cite{lan2019albert}. Nevertheless, these VC frameworks are only able to train small models that can fit entirely in any volunteer device. Petal~\cite{borzunov2022petals} and Swarm~\cite{ryabinin2021swarm} seek to extend pipeline parallelism to a large number of heterogeneous volunteer servers. While they focus on fault tolerance in distributed training, it does not adequately address the low network bandwidth and high latency in a decentralized network, thereby unable to efficiently perform LLM training.

\subsection{Distributed Training}

Distributed training leverages multiple GPUs to expedite training or manage large models. \textit{Data parallelism}~\cite{sergeev2018horovod, jiang2020unified, moritz2018ray} divides the dataset among GPUs, with each GPU training a model replica on a unique data batch. Gradients are then shared to synchronize model parameters. However, this method doesn't cater to large model sizes. Alternatively, \textit{model parallelism} splits the model into sub-models for different GPUs. While this reduces memory needs per GPU, it has drawbacks: only one GPU operates at a time due to sub-model dependencies, and significant inter-GPU communication is required, especially with intra-layer partitioning~\cite{shazeer2018mesh, lepikhin2020gshard, xu2021gspmd, athlur2022varuna}.

\begin{figure}[ptb!]
\setlength{\belowcaptionskip}{-0.4cm}
\centering
\includegraphics[width=0.8 \columnwidth]{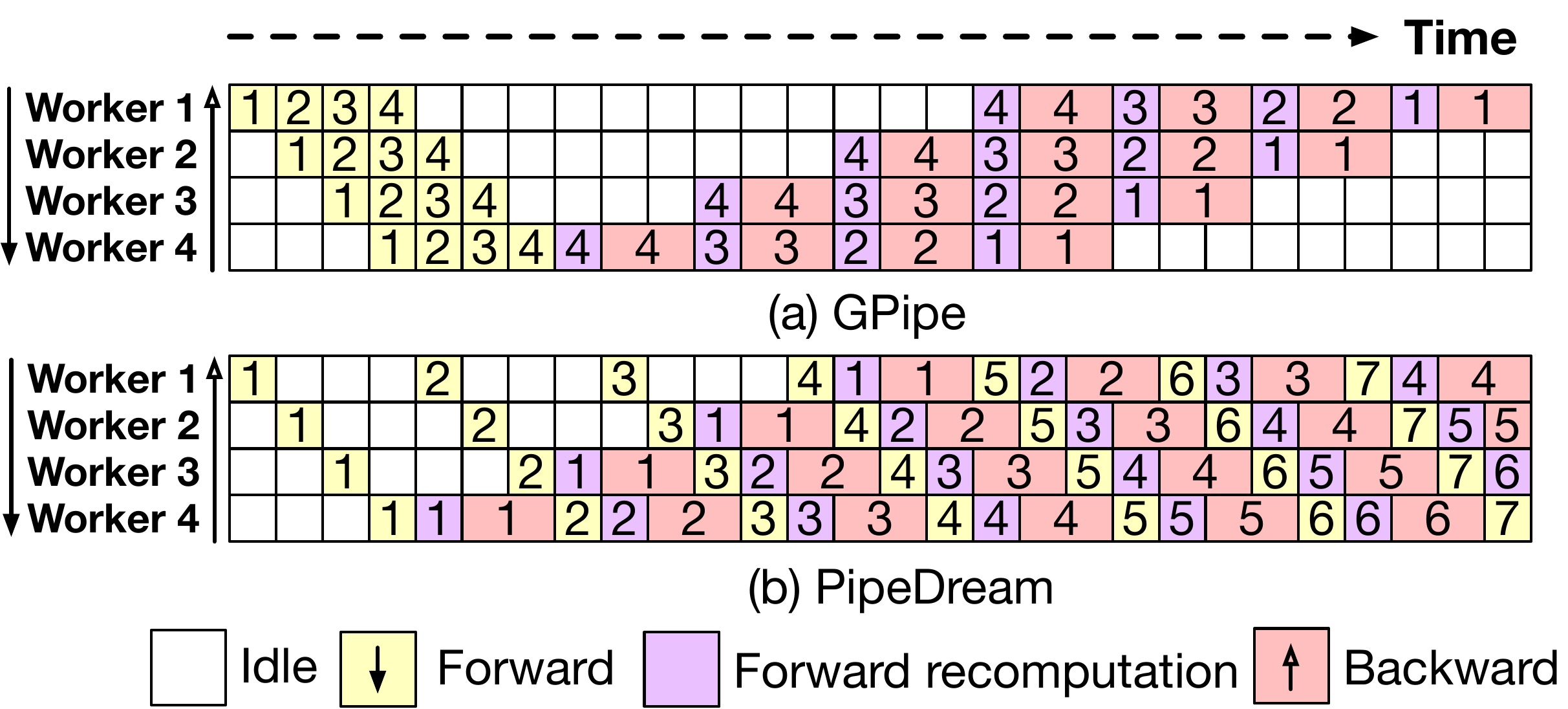}
\caption{GPipe and PipeDream.}
\label{figs:gpipe_pipedream}
\end{figure}

\noindent \textbf{Pipeline parallelism} addresses the GPU utilization issue in model parallelism. It divides the model similarly but processes different mini-batches concurrently on separate GPUs. In \textit{synchronous training}, all GPUs train in lockstep, while in \textit{asynchronous training}, each GPU trains independently and periodically updates a shared model. Synchronous training typically converges faster but has lower throughput and GPU utilization. Figure~\ref{figs:gpipe_pipedream} illustrates two pipeline parallelism methods: GPipe~\cite{huang2019gpipe} and PipeDream~\cite{narayanan2019pipedream}. GPipe uses synchronous training, leading to GPU idleness between forward and backward propagation. PipeDream, on the other hand, employs asynchronous training, maximizing GPU utilization but potentially slowing convergence. FTPipe~\cite{eliad2021fine} is another asynchronous method that allows more flexible layer placements on GPUs, demanding a faster interconnect.

\noindent \textbf{GPU memory optimization}. Training large models like GPT-3 demands extensive GPU resources. For instance, GPT-3's parameters alone require 700 GB (in FP32). With optimizer states typically being 3x the model size, the total memory exceeds 2.8 TB, making single-GPU training impossible. To address this, memory-saving techniques have been proposed. The ZeRO (Zero Redundancy Optimizer) family~\cite{rajbhandari2020zero, ZeRO-Offload, rajbhandari2021zero} under the DeepSpeed~\cite{deepspeed} framework optimizes memory usage. ZeRO and FSDP~\cite{fsdp} distribute model states across GPUs, fetching them on-demand, albeit with increased communication overhead. ZeRO-Offload~\cite{ZeRO-Offload} transfers intermediate training data to host memory, while ZeRO-Infinity~\cite{rajbhandari2021zero} utilizes CPU, GPU, and NVMe memory for larger model training. Techniques like forward recomputation, used in GPipe and PipeDream, also help by recalculating activation outputs during backward propagation instead of storing them.

\noindent \textbf{Limitations}.
{\em First}, the existing distributed training schemes require high-speed interconnect, such as NVLink~\cite{NVLink} and Infiniband, to effectively construct a training pipeline that overlaps sub-model computation with the transmission model states between sub-models (GPUs). However, advanced network devices are as scarce and expensive as high-end accelerators, rarely accessible to average users. {\em Second}, there is a lack of fault tolerance in the design of distributed training and failures at any workers will stall the entire training pipeline. {\em Third}, a model must be statically partitioned offline before execution. Any changes to the number of workers or the network condition requires the model to be re-partitioned, lacking flexibility and elasticity.   

\section{\atom}

\subsection{Design Overview}

Unlike traditional methods that distribute sub-models across multiple GPUs, \atom houses the entire model in a single server's host memory. When required, it employs a memory swapping technique to transfer model portions to the GPU. Notably, \atom's use of memory swapping in LLMs distributed training is pioneering, even though the concept is not new. Solutions like ZeRO-Offload~\cite{ZeRO-Offload} also use memory swapping but not explicitly for distributed training. Furthermore, \atom adopts asynchronous training, avoiding the lock step constraints seen in platforms like DeepSpeed. While DeepSpeed merges device memory across GPUs, leading to communication overheads, \atom efficiently uses host memory, making it a resource-savvy choice in a decentralized environment. Besides, \atom does not have a single point of failure and is able to make training progress in the presence of worker failures, joining, and departure. 

In the subsequent sections, we delve into the analytical findings of the GPT-3 model, which served as a catalyst for the \atom architecture (\S~\ref{sec:gpt-3}). We further discuss the intricacies of sub-model swapping, which ensures optimal GPU utilization (\S~\ref{sec:swap-memory}), followed by an exploration of the automated model segmentation and code synthesis processes (\S~\ref{sec:model-partitioning}).

\subsection{Characterization of GPT-3}
\label{sec:gpt-3}
We study the feasibility of partitioning GPT-3 to fit into a single GPU and streamlining sub-model execution and switching. We perform a comprehensive profiling of GPT-3 to measure the memory footprint, the execution time, and the memory swapping time of its key layers and associated data during training. The full version of GPT-3 175B includes a chain of 96 identical decoders and requires approximately 2.8 TB memory. To fit GPT-3 in our server, which is configured with 384 GB host memory, we trimmed GPT-3 175B down to including two decoders. Note that the trimmed GPT-3 model has identical structure in model execution, including the profiling of one of the 96 identical decoders, but differs in the accuracy of the trained model. 

We constructed the GPT-3\footnote{\url{https://github.com/karpathy/minGPT}} computation graph through PyTorch module {\tt torch.nn.Module} and enabled tracing. Layers and operators were executed according to the graph's topological order. All results were the average of 10 runs. The experiments were conducted on an NVIDIA Tesla V100 GPU with 32 GB device memory running on a PCI-e 3.0 bus.

\begin{figure}[ptb!]
\setlength{\belowcaptionskip}{-0.4cm}
\centering
\includegraphics[width=0.7 \columnwidth]{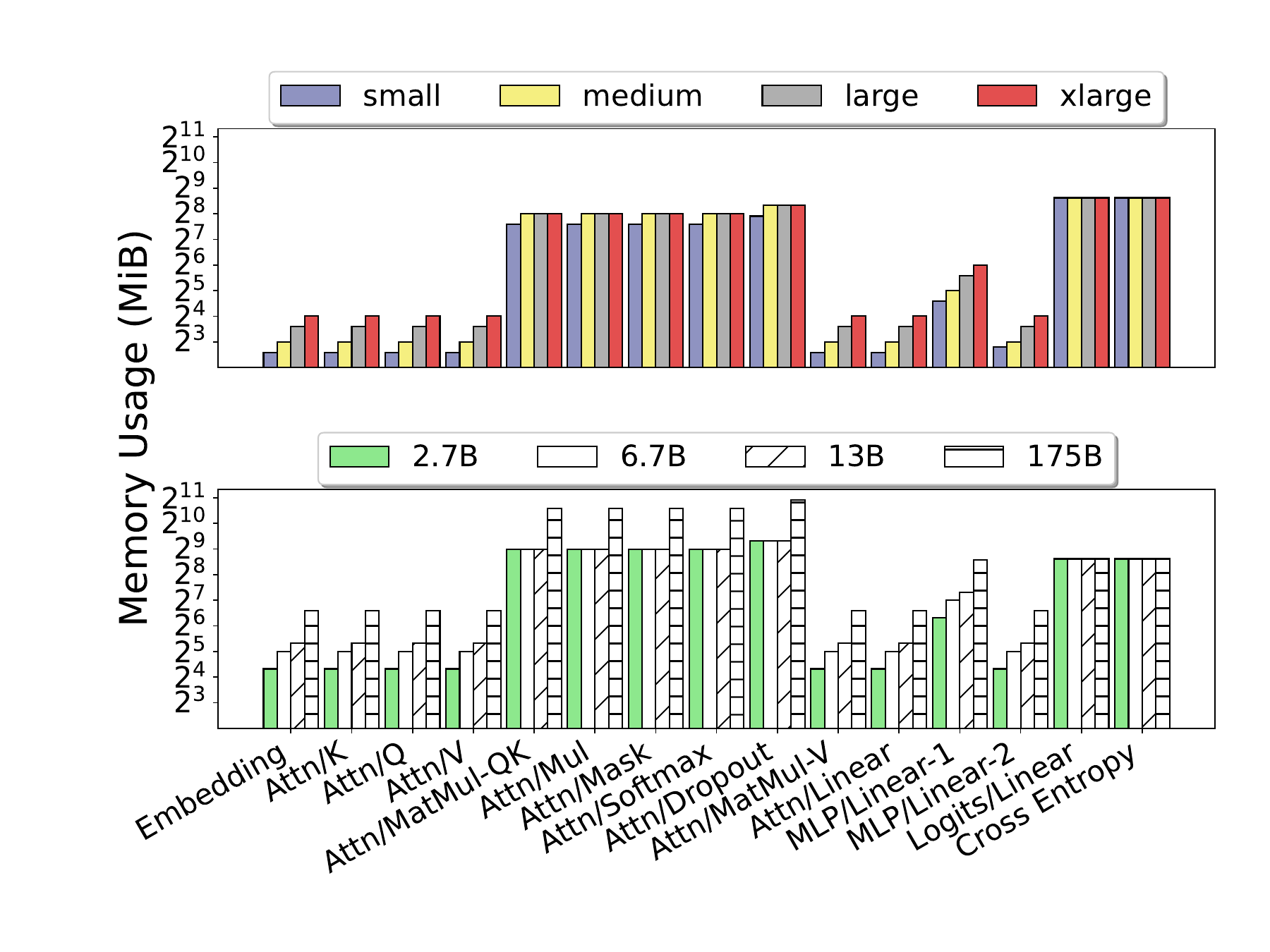}
\caption{GPT-3 memory usage in forward propagation.}
\label{figs:gpt3_working_memory_usage_fw}
\end{figure}

\subsubsection{Memory usage breakdown}

\begin{figure}[ptb!]
\setlength{\belowcaptionskip}{-0.4cm} 
\centering
\includegraphics[width=0.7 \columnwidth]{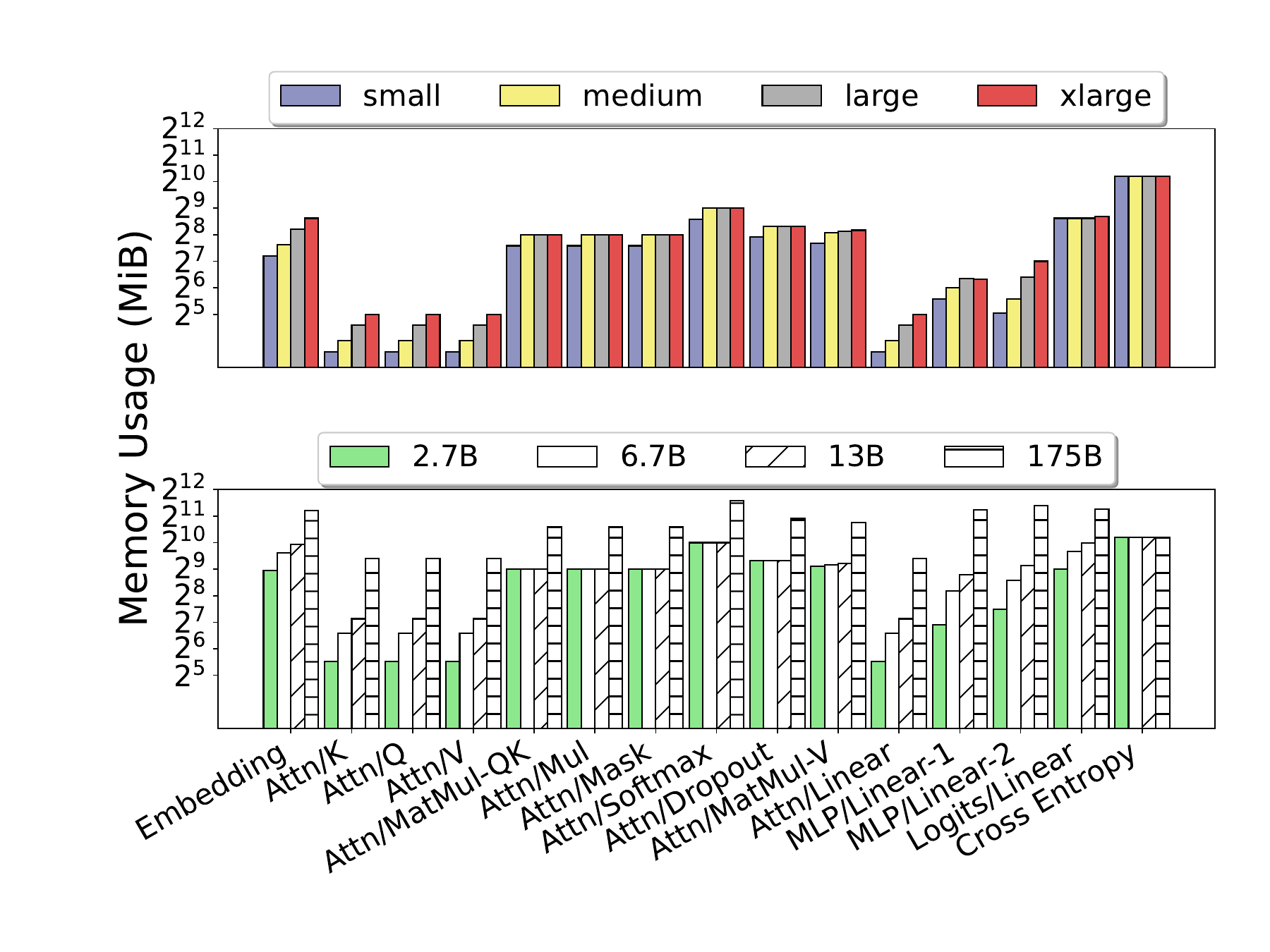}
\caption{GPT-3 memory usage in backward propagation.}
\label{figs:gpt3_working_memory_usage_bw}
\end{figure}

Training GPT-3 175B model is extremely memory-consuming. This section characterizes the breakdown of GPU memory consumption to train a GPT-3 model. 
Figure~\ref{figs:gpt3_working_memory_usage_fw} and Figure~\ref{figs:gpt3_working_memory_usage_bw} show the memory footprints of individual layers in GPT-3 in forward propagation and backward propagation, respectively. We show eight variants of GPT-3 ranging from 125 million (small) to 175 billion (175B) parameters. 
Memory usage was measured by \texttt{reset\_peak}, \texttt{\_memory\_stats}, and \texttt{max\_memory\_allocated} from the PyTorch library \texttt{torch.cuda}. The measured memory consumption of each layer includes activation inputs, intermediate outputs, model parameters, and memory for temporary workspace at runtime. 

As shown in the figures, the peak memory usage in GPT-3 175B are the {\em dropout} layer (1,920 MB) in forward propagation and the {\em softmax} layer (3,072 MB) in backward propagation. The results suggest that even the most memory-intensive layer in the GPT-3 175B model can fit in an entry-level GPU, e.g., NVIDIA GTX 1080 with 8 GB device memory, without a need to further perform intra-layer partitioning~\cite{jia2018exploring, lepikhin2020gshard, eliad2021fine}. 




\begin{table}[h]
\setlength\tabcolsep{0pt} 
\caption{GPT-3 configurations to measure the cost of network transmissions in model partitioning.}
\label{table:gpt3-config}
\begin{tabular*}{\columnwidth}{@{\extracolsep{\fill}} l *{5}{c}}
\toprule[1.5pt]
\multicolumn{1}{c}{GPT-3 Model} & 
\multicolumn{1}{c}{$L$} & 
\multicolumn{1}{c}{$d_{model}$} &  
\multicolumn{1}{c}{$n_{heads}$} & 
\multicolumn{1}{c}{activation payload (MiB)} \\
\midrule
Small (125M)        & 2048 & 768   & 12 & 6 \\
Medium (350M)       & 2048 & 1024  & 16 & 8 \\
Large (760M)        & 2048 & 1536  & 16 & 12 \\
XL (1.3B)           & 2048 & 2048  & 24 & 16 \\
2.7B                & 2048 & 2560  & 32 & 20 \\
6.7B                & 2048 & 4096  & 32 & 32 \\
13B                 & 2048 & 5120  & 40 & 40 \\
175B                & 2048 & 12288 & 96 & 96 \\
\bottomrule[1.5pt]
\end{tabular*}
\end{table}

\subsubsection{Memory swapping vs. network transmission} \label{sec:swap-preload-bandwidth}

Having demonstrated the capability to accommodate a GPT-3 model on a singular GPU, our subsequent investigation focuses on the comparative efficiency of executing Large Language Models (LLMs) on a single GPU versus distributed GPUs across a network. To maintain the integrity of our comparison, we ensure uniformity in both the quantity and specifications of GPUs utilized during the training phase. The \atom mechanism deploys full replicas of models on distinct GPUs through sub-model swapping, leveraging the allreduce communication protocol for model synchronization. In contrast, pipeline parallelism techniques partition the model across multiple GPUs, necessitating network transmissions to relay activation outputs between sub-model segments.

In our endeavor to quantify the overhead associated with transmitting activation and intermediate tensors between sub-models across a network, we established a distributed training environment using PyTorch on a pair of servers. This was facilitated by the asynchronous gRPC APIs~\cite{grpc}, as implemented in Hivemind~\cite{hivemind}. Our interconnection medium was a 10 Gbps Ethernet. To comprehensively evaluate transmission rates, we employed bandwidth throttling at 400 Mbps, 800 Mbps, and the maximum 10 Gbps. It's noteworthy that the 400 Mbps rate mirrors the standard wide-area bandwidth observed in decentralized data analytics~\cite{wang2018dynamic}. For the purpose of establishing an upper limit on achievable bandwidth, we utilized the loopback network device, denoted as {\em localhost}. We meticulously partitioned the GPT-3 models at the boundaries of the transformer blocks (decoders), given that this location exhibits the minimal activation tensor size. This strategic partitioning provides a lower bound on network transmission overhead for any model segmentation. Table \autoref{table:gpt3-config} delineates the configuration specifics of GPT-3 and the resultant activation payload necessitating network transmission. The activation tensor adopts the shape $[batch_size, sentence_length, d_model]$, with a batch size of 1, an input sequence length $L$ of 2048, and an embedding dimension represented by $d_model$.

\begin{figure}[ptb!]
\setlength{\abovecaptionskip}{-0.1cm}
\setlength{\belowcaptionskip}{-0.1cm}
\centering
\includegraphics[width=0.65 \columnwidth]{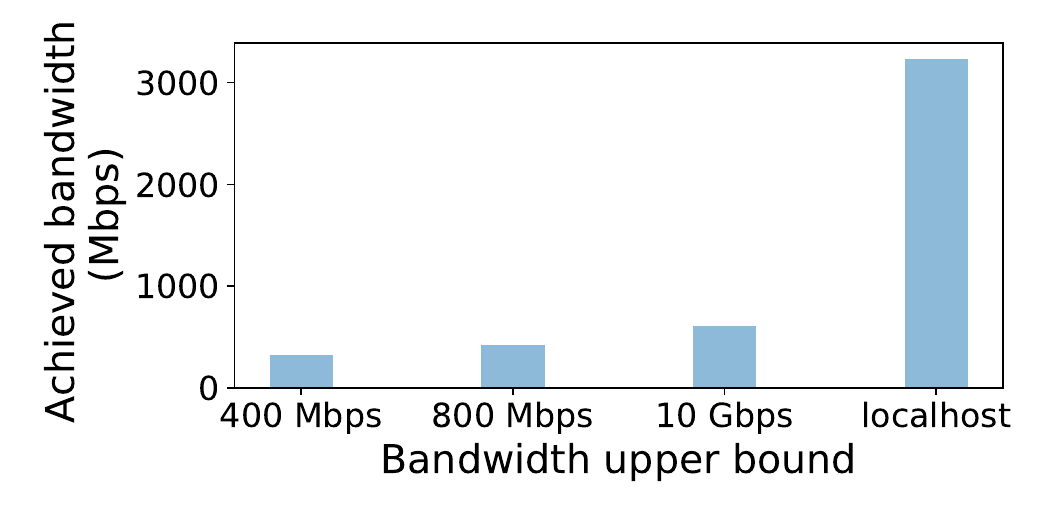}
\caption{Achievable bandwidth in distributed training.}
\label{figs:bandwidth}
\end{figure}


\begin{figure}[ptb!]
\centering
\includegraphics[width=0.6 \columnwidth]{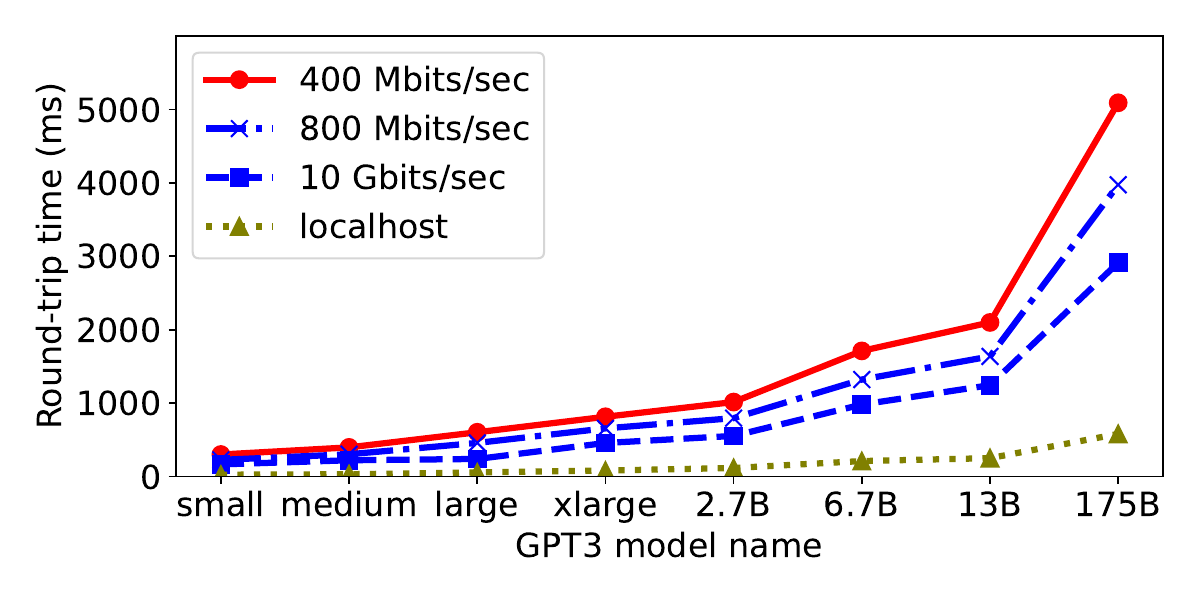}
\caption{Transmission time of activation payload using gRPC in various GPT-3 models.}
\label{figs:round-trip-time}
\end{figure}

\autoref{figs:bandwidth} illustrates the bandwidth achieved when transmitting activation payloads between sub-models under diverse network conditions. In the context of a 10 Gbps network, the transmission of activations is constrained to a peak of $610$ Mbps, a phenomenon also documented by~\cite{biswas2018designing}\footnote{This pertains to a round-trip required to invoke the functions of remote sub-models}. While gRPC is a prevalent protocol in the realm of distributed training, it mandates that activation payloads be relayed from the GPU to the CPU for serialization prior to their dispatch to a subsequent sub-model. \autoref{figs:round-trip-time} presents the cumulative transmission duration, encompassing the journey from the originating GPU to the CPU and subsequently to the GPU of the next sub-model, for activation payloads across various GPT-3 model configurations.

\begin{figure}[ptb!]
\setlength{\belowcaptionskip}{-0.2cm}            
\centering
\includegraphics[width=0.7 \columnwidth]{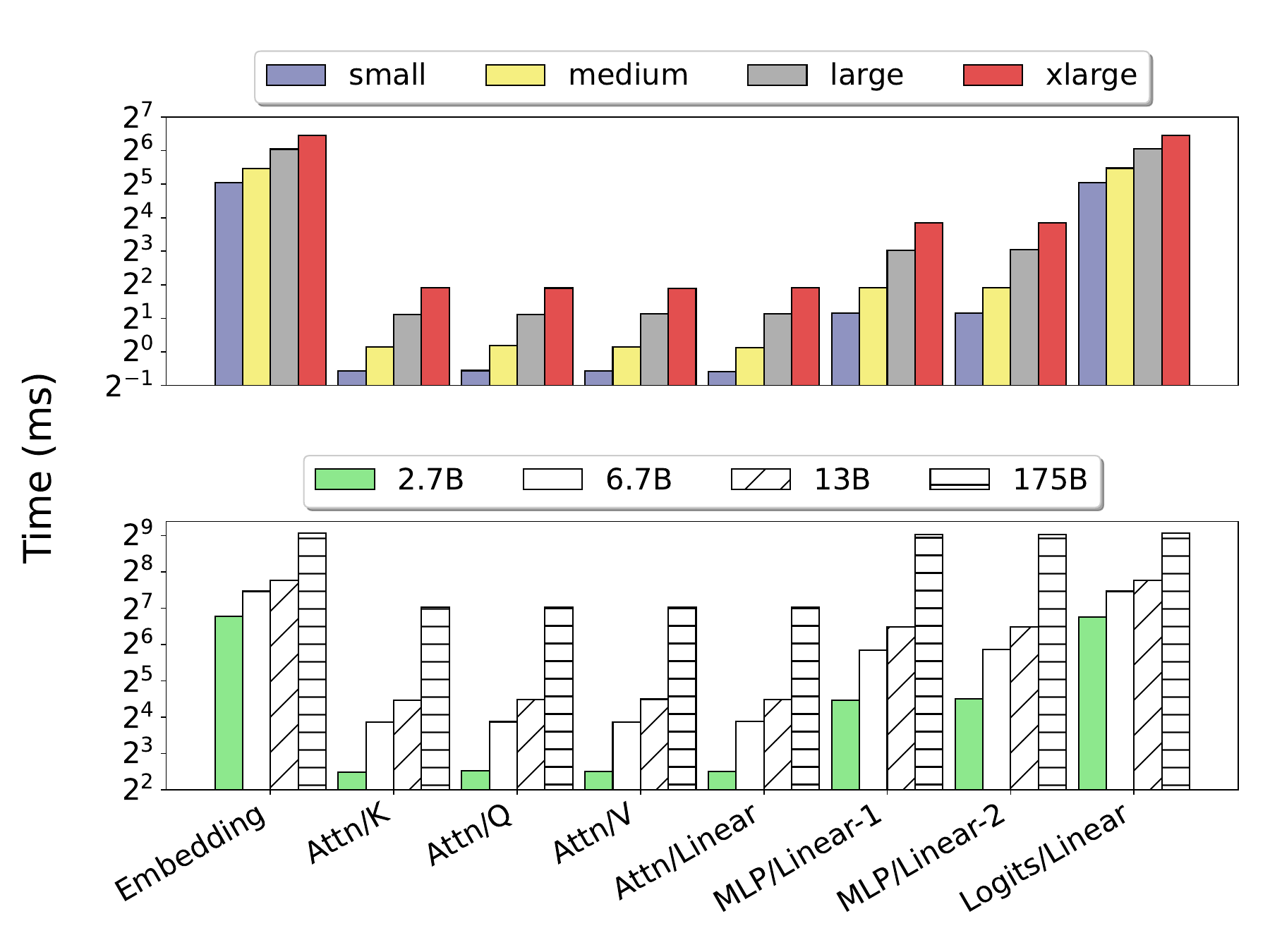}
\caption{Layer loading time in various GPT-3 models.}
\label{figs:gpt3_model_loading_node_times}
\end{figure}

\begin{figure}[ptb!]
\centering
\includegraphics[width=0.65 \columnwidth]{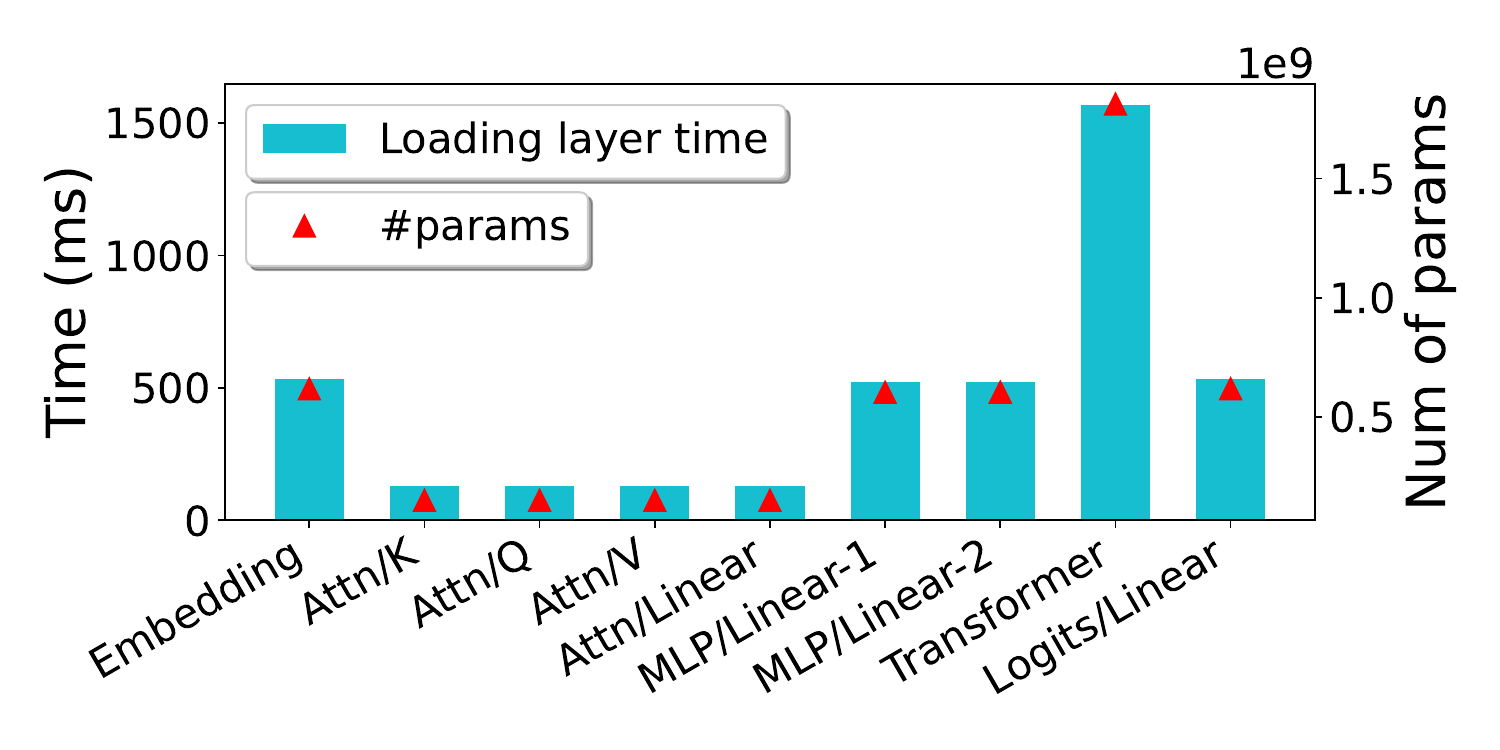}
\caption{Relationship between loading time and layer size.}
\label{figs:gpt3-175b-model-loading-node-times-and-num-params}
\end{figure}

We evaluated the time required to load key GPT-3 model layers from host to GPU memory through memory swapping. This layer loading time, akin to activation transmission in model and pipeline parallelism, critically impacts training efficiency and GPU utilization. \autoref{figs:gpt3_model_loading_node_times} depicts this for eight GPT-3 models, revealing that even for the largest GPT-3 175B model, layer loading is substantially faster than the activation transmission time shown in \autoref{figs:round-trip-time}. This transmission time, resulting from optimal transformer block partitioning, could increase with different partitioning strategies. While activation tensor size varies based on output type and partitioning, layer loading time is consistent, scaling with the number of model parameters. \autoref{figs:gpt3-175b-model-loading-node-times-and-num-params} confirms this linear trend. Notably, loading the {\em Logits} layer post the {\em Transformer} block is about 6$\times$ faster than transmitting its activation output over a 10 Gbps network. This disparity will grow in decentralized, wide-area network settings, and with the upcoming PCI-e 5.0, the gap is poised to expand further.

\subsection{Streamlining Sub-model Swapping}~\label{sec:swap-memory}

\begin{figure}[ptb!]
\centering
\includegraphics[width=0.7 \columnwidth]{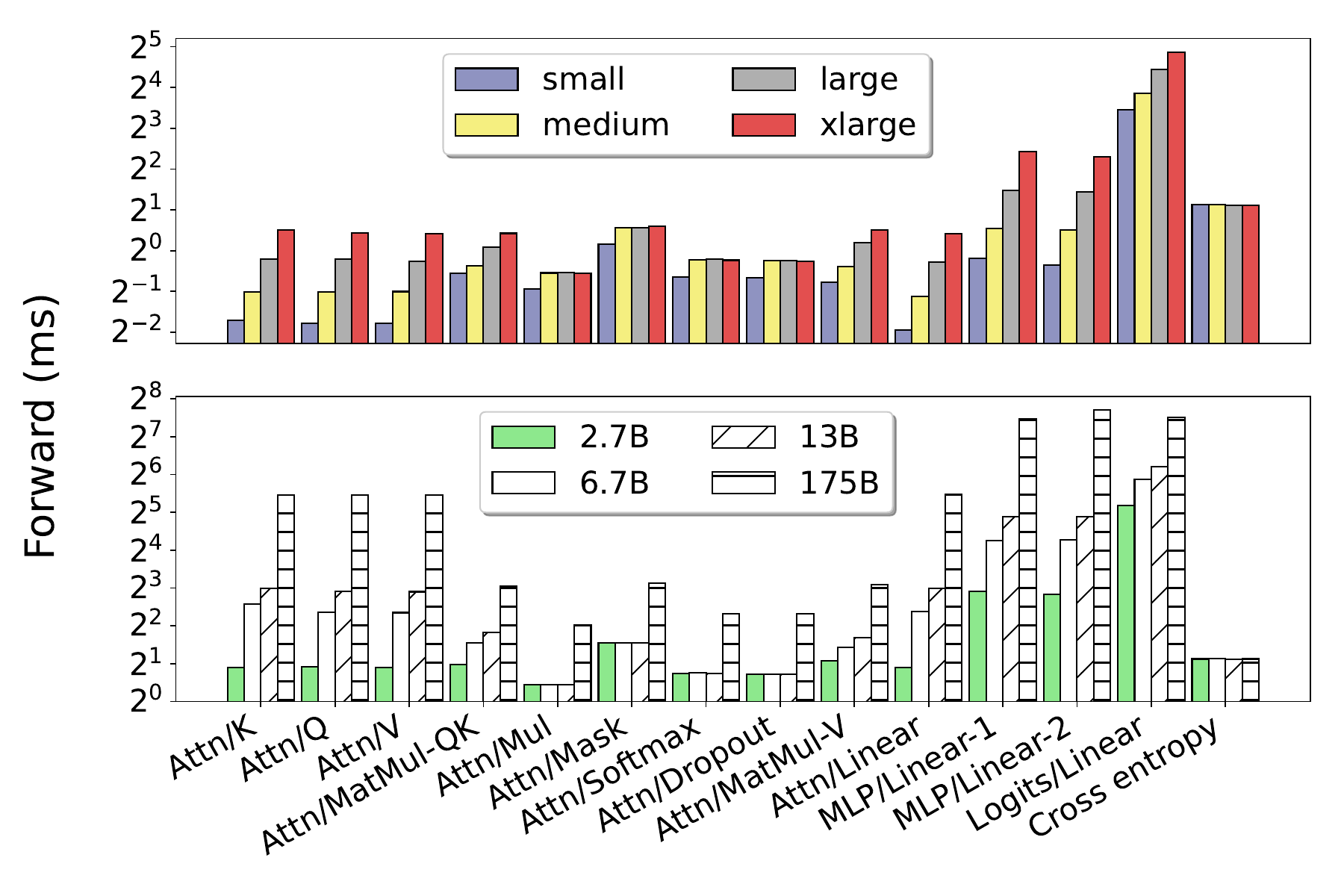}
\caption{Breakdown of forward propagation time in GPT-3.}
\label{figs:fw-time}
\end{figure}
\begin{figure}[ptb!]
\centering
\includegraphics[width=0.7 \columnwidth]{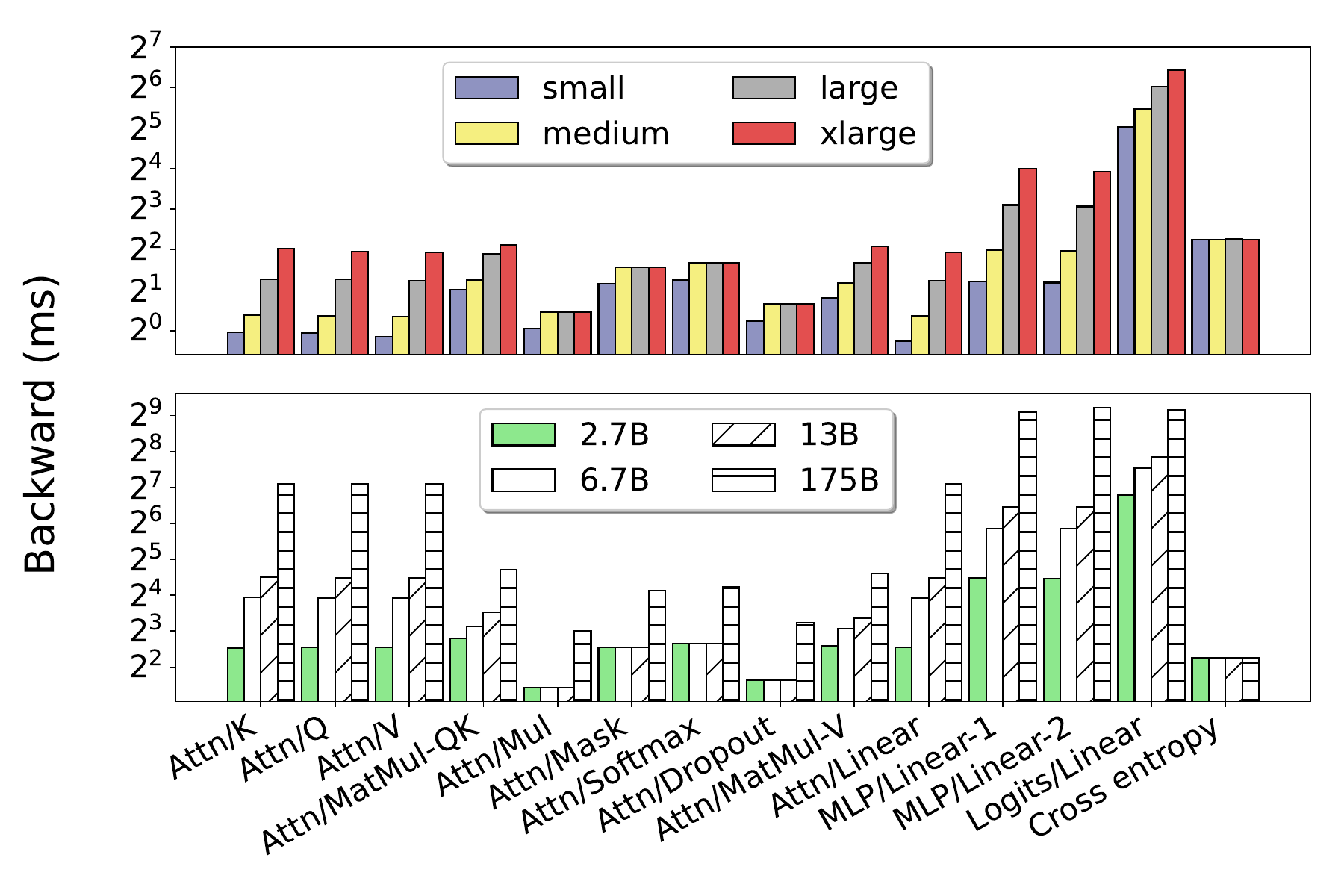}
\caption{Breakdown of backward propagation time in GPT-3.}
\label{figs:bw-time}
\end{figure}

Our profiling underscores the superiority of memory swapping over network transmission in terms of efficiency. However, orchestrating a model swapping schedule that concurrently manages model execution and loading remains intricate. \autoref{figs:fw-time} and \autoref{figs:bw-time} show the layer-by-layer breakdown of forward and backward propagation durations, respectively. Intriguingly, the durations for backward propagation align closely with the loading times presented in Figure~\ref{figs:gpt3_model_loading_node_times}. This congruence presents a potential avenue to synchronize sub-model execution with its loading during the backward phase. 

A prominent challenge emerges from the observation that forward propagation durations are invariably shorter than their corresponding layer loading times, leading to execution pauses as the system awaits the next sub-model's readiness. To mitigate this, \atom employs a strategy of extending forward propagation durations by processing multiple mini-batches consecutively before initiating the backward phase. This gradient accumulation approach, a staple in asynchronous training, ensures a favorable computation-to-swapping balance during forward propagation.

\begin{figure}[ptb!]
\setlength{\belowcaptionskip}{-0.2cm}
\centering
\includegraphics[width=0.75 \columnwidth]{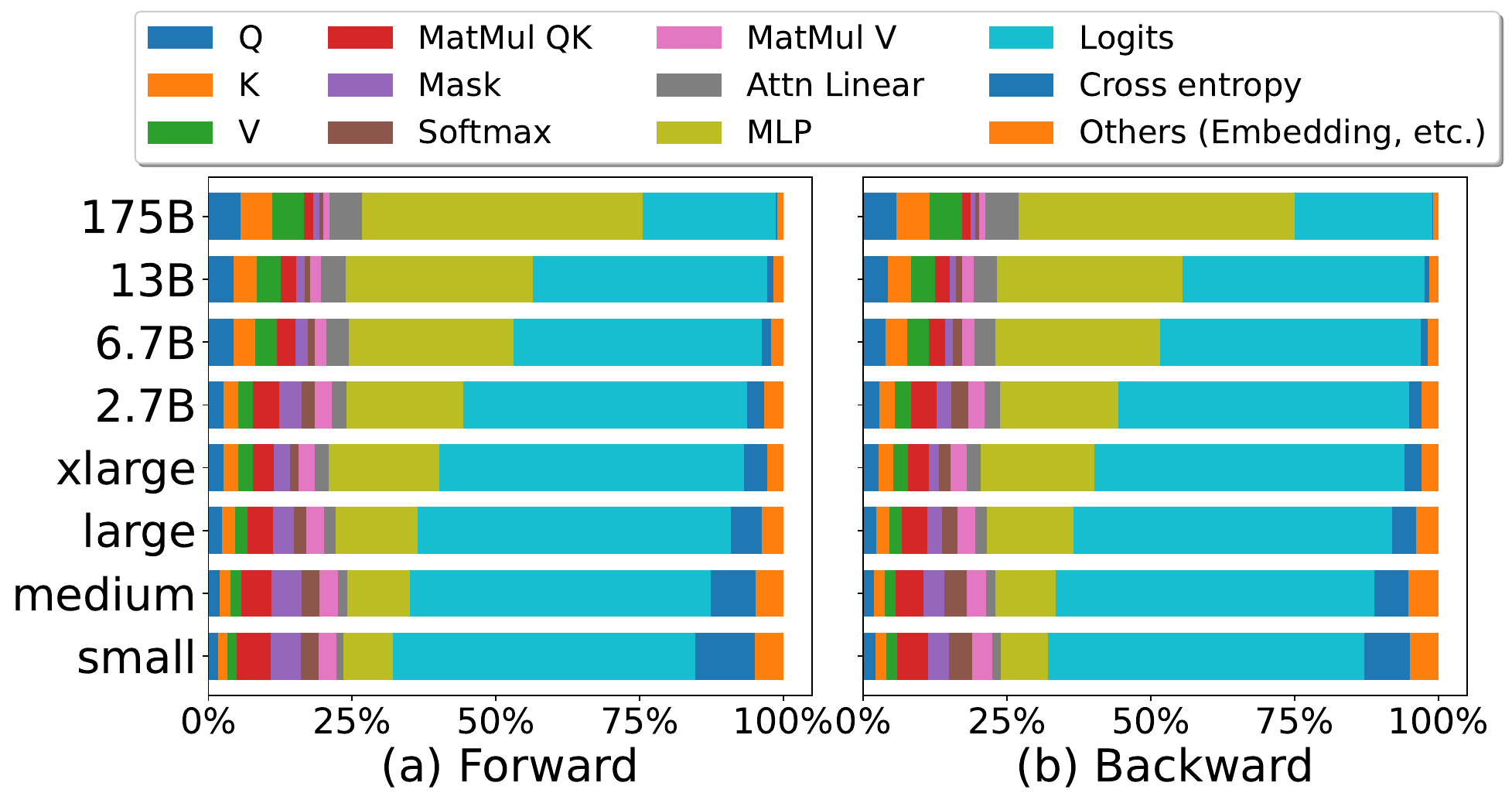}
\caption{Time breakdown of operators in a transformer block, including the execution time on the {\em Embedding} layer. GPT3-175B a reduced size of 2-layer subset.}
\label{figs:fw-bw-percentage}
\end{figure}

Another particularly intricate issue pertains to layers characterized by substantial loading durations but minimal computational overhead. The {\em embedding} layer serves as a prime exemplar. While its size mandates a considerable loading time, as evidenced in \autoref{figs:gpt3_model_loading_node_times}, its computational demands are minimal. \autoref{figs:fw-bw-percentage} elucidates the relative computational time of the embedding layer within a transformer block. As the figure delineates, the computational time of a embedding layer is marginal during both forward and backward propagations, and this proportion diminishes with escalating model sizes. However, persistently caching this layer on a GPU amplifies memory consumption. Conversely, relegating the embedding computation to the CPU, especially for sizable embeddings (approximating 2.4 GB with a 50K vocabulary and d\_model=12288 in FP32 format), on GPUs with limited resources can inadvertently compromise performance.

\begin{figure}[ptb!]
\setlength{\belowcaptionskip}{-0.2cm}
\centering 
\includegraphics[width=0.7 \columnwidth]{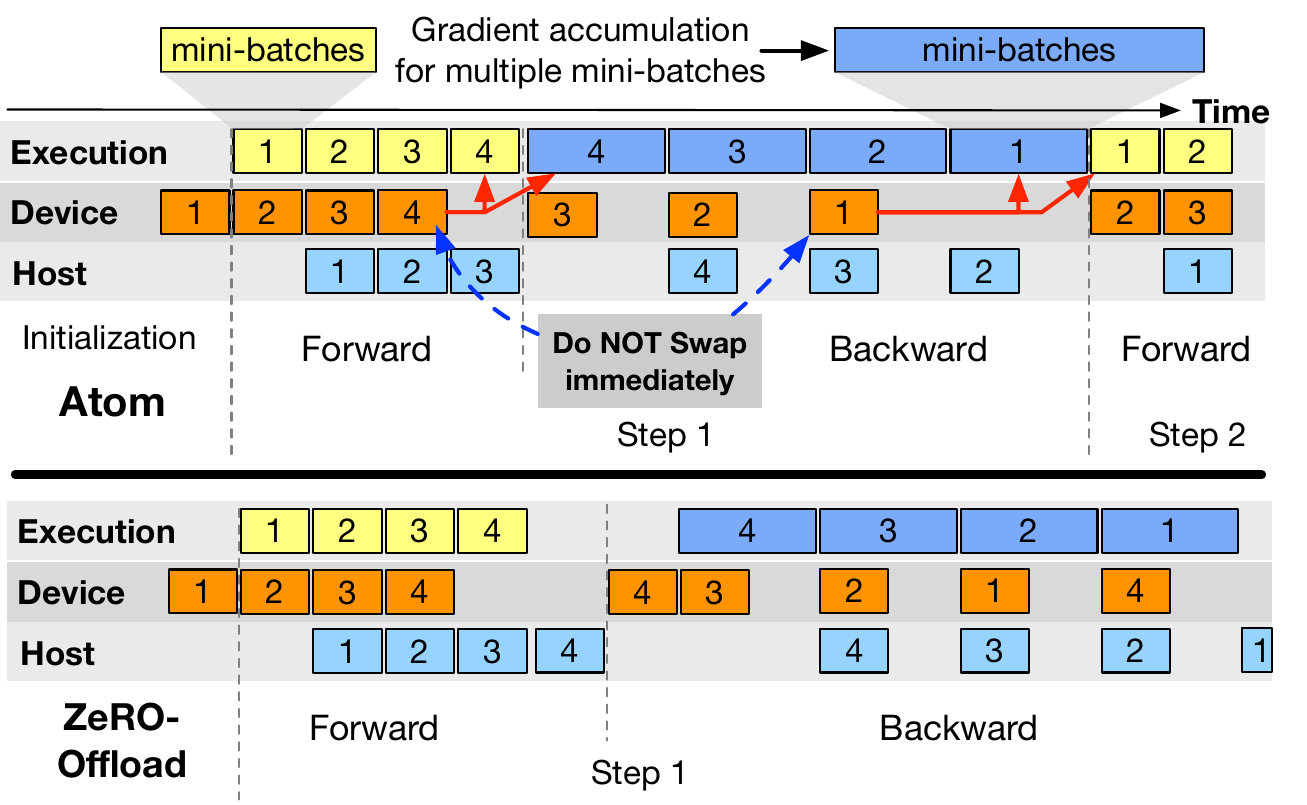}
\caption{Schedule of sub-model swapping. Sub-models are denoted by numbered blocks. The {\em embedding} layer is included in sub-model 1. \atom demonstrates improved efficiency compared to the ZeRO-Offload schedule.}
\label{figs:swap}
\end{figure}

\autoref{figs:swap} presents the scheduling intricacies of sub-model swapping within the \atom framework. The dotted vertical demarcations distinguish individual training iterations, each encompassing a forward and a backward pass. The {\em execution} trajectory delineates the temporal progression of sub-model operations, while the {\em device} and {\em host} trajectories indicate the temporal residence of sub-models in device and host memories, respectively.

A salient observation from \autoref{figs:swap} is the inherent locality advantage in LLM training with model swapping. Specifically, the terminal sub-model of the forward pass is instantaneously requisitioned for the subsequent backward pass. Similarly, the concluding sub-model of the backward phase becomes the precursor for the ensuing training iteration's forward pass. Given that the embedding layer is encapsulated within sub-model 1 and remains undisturbed post the backward phase, it stands primed for immediate deployment as the next forward iteration commences. This strategic retention obviates potential GPU dormancy stemming from the temporal disparity between execution and loading times inherent to the embedding layer.

Furthermore, \autoref{figs:swap} underscores the efficacy of gradient accumulation across multiple mini-batches. This strategy ensures that the computational durations of layers, during both forward and backward phases, consistently surpass their respective loading times, thereby circumventing any GPU inactivity.

\begin{figure}[ptb!]
\setlength{\belowcaptionskip}{-0.2cm}
\centering
\includegraphics[width=0.65 \columnwidth]{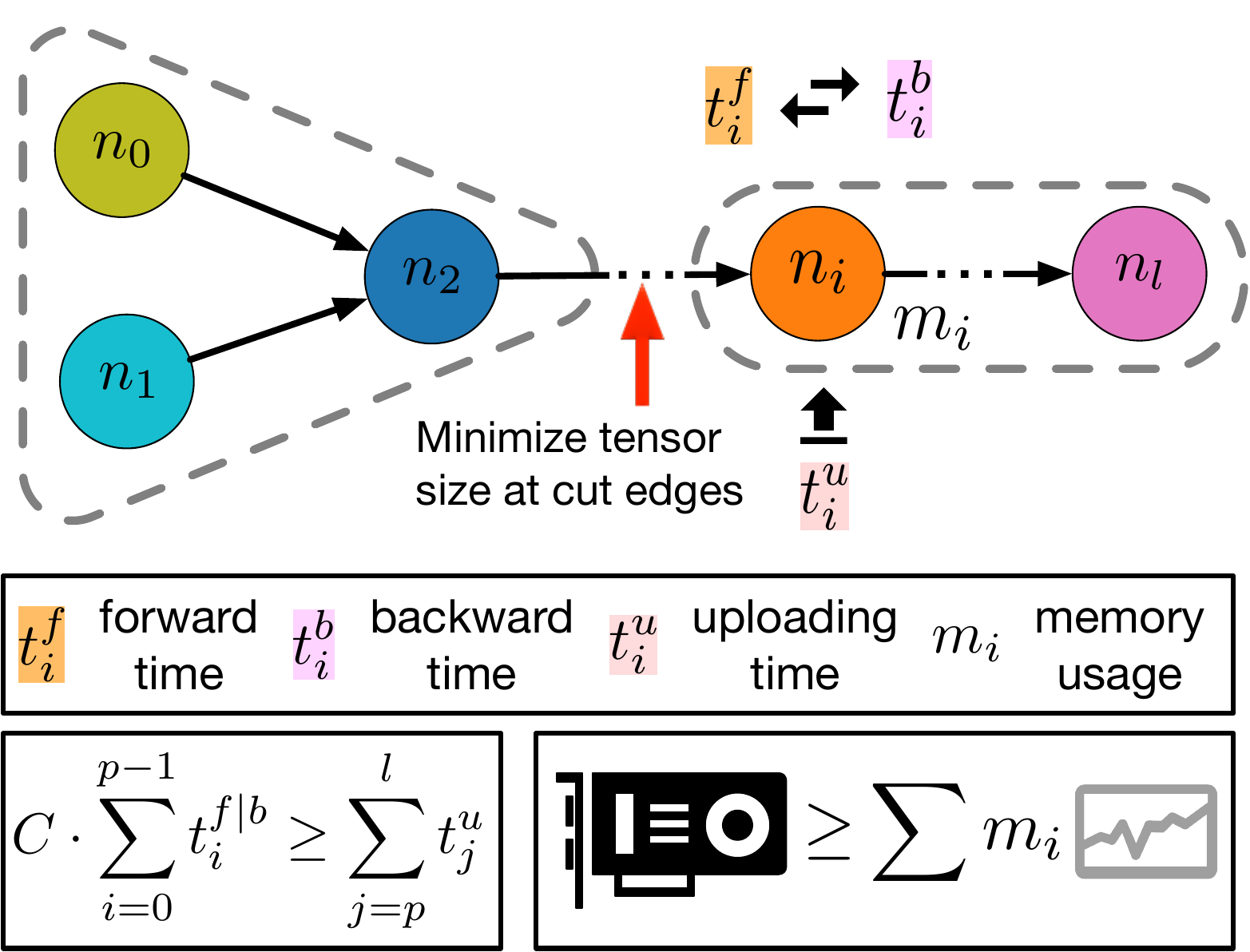}
\caption{The formulation of model partitioning.}
\label{figs:partitioning}
\end{figure}

\subsection{Model Partitioning and Code Generation} ~\label{sec:model-partitioning}
The key to build a model swapping schedule that seamless overlaps model execution and loading is to determine a model partitioning scheme that satisfies that 1) any sub-model fits in the GPU device memory; 2) adjacent sub-models in the topological order in the model's computation graph have comparable execution and loading time. To guide model partitioning, we augment the computation graph with operator/layer's forward/backward execution time, loading time, and memory usage. To collect each operator's information, the model is profiled offline, in which the computation graph is executed operator (layer) by operator (layer). Profiled operators are swapped out to host memory to ensure that even the largest GPT-3 175B model can be profiled on a single GPU. We devise an heuristic exhaustive search algorithm to find a valid partitioning that satisfies all constraints.

\noindent \textbf{Problem definition.} A neural network can be represented as a directed computation graph (DAG) $G(E, V)$, where each node $v \in V$ is a layer or an operator, and an edge $e \in E$ is an intermediate input or output tensor to the neighboring nodes. The graph partitioning problem can be expressed as an optimization problem subject to multiple constraints, as shown in Figure~\ref{figs:partitioning}. We design an algorithm to automatically partition the model in \atom subject to constraints, including the computation-to-loading ratio between adjacent sub-models, GPU memory capacity, the minimization of tensor size at the cutting of edges. Node ID $i$ is sorted in the topological order of the computation graph and indexed from $0$ to $l$. Attribute information of each node includes the max working memory size $m_i$ at runtime, the forward execution time $t_i^f$, the backward execution time $t_i^b$, and the loading time $t_i^u$ from host memory to device memory. 

\begin{algorithm}
\caption{Partitioning a GPT-3 model.}\label{alg:partition}
\begin{algorithmic}[1]

\Function{ValidConstraints}{$G,c_s,c_e,l_s,l_e$}{
    \If{$G.mem(c_s,c_e) \leq GPU\_capacity \AND \textcolor{olive}{// pruning} \newline 
        \hspace*{2.5em} G.mem(l_s,l_e) \leq GPU\_capacity  \AND \newline 
        \hspace*{2.5em} G.comp\_t(c_s,c_e) \geq G.load\_t(l_s,l_e) $ 
    }
        \State \textbf{return True}
    \Else{
        \State \textbf{return False}
    }
    \EndIf
}
\EndFunction

\State \textcolor{olive}{// $c_s$ and $c_e$ are the begin and end operator indexes considered for sub-model execution; $l_s$ and $l_e$ are operator indexes for sub-model preloading; the objective is to match the computation time between $c_s$ and $c_e$ with the loading time between $l_s$ and $l_e$.}
\Function{PartitionModel}{$G,c_s,c_e,l_s,l_e,t,partitions$}{
    \If{!$ValidConstraints(G,c_s,c_e,l_s,l_e)$}
        \State \textbf{return}
    \EndIf

    \If{$l_e = G.num\_nodes -1$ \AND len($t$) $\geq$ 0}
        \State $partitions$.insert($t$)
        \State \textbf{return}
    \EndIf
    \State \textcolor{olive}{// squeeze boundary to keep more nodes within a certain range}
    \For{$\hat{l_e} \gets (G.num\_nodes-1$ to $l_e$, $step\_size)$}
        \State $t$.insert($(l_s,l_e),(l_e+1,\hat{l_e})$)
        \State \textcolor{olive}{// recursively search for a valid partition}
        \State $PartitionModel(G,l_s,l_e,l_e+1,\hat{l_e},t,partitions)$
        \State $t$.pop($(l_s,l_e),(l_e+1,\hat{l_e})$) \textcolor{olive}{// backtracking}
    \EndFor
}
\EndFunction

\Function{\textcolor{red}{Main}}{$G$}{ \textcolor{olive}{// graph G is topologically sorted} 
    \State $c_s, t, partitions \gets 0, [], []$  \textcolor{olive}{// partial and final results}
    \State \textcolor{olive}{// search valid partitions of the computation graph}
    \For{$c_e \gets G.num\_nodes-2$ to $c_s$; $l_s \gets c_e+1$} 
        \For{$l_e \gets (G.num\_nodes$ to $l_s, step\_size)$}
            \State $PartitionModel(G,c_s,c_e,l_s,l_e,t,partitions)$
        \EndFor
    \EndFor
}
\EndFunction
\end{algorithmic}
\end{algorithm}


The \textit{objective function} is to minimize the computation time, i.e., the sum of all sub-model forward and backward propagation time, and to minimize the partitioning cost, i.e., the total size of intermediate tensors (the size of cutting edge) across different partitions. 
One \textit{constraint} is GPU memory capacity. The mainstream commodity GPU memory size is 11 GiB, 16 GiB, or 32 GiB. Another constraint the computation/prelading ratio where the partitioning result should ensure that the current sub-model computation time can overlap the next sub-model layer preloading time. 
Given a bipartite model partitioning ${\{n_0, \dots, n_i\}}$ and ${\{n_{i+1}, \dots, n_l\}}$, it should satisfy the condition that both $\sum_{i=0}^{p} m_i$ and $\sum_{k=i+1}^{l} m_k$ do not exceed the GPU memory capacity. Besides, computation can overlap with next sub-model loading $C \cdot \sum_{k=0}^{i-1} t^{f|b}_{k} \geq \sum_{k=i}^{l} t^{o}_{k}$, where $C$ is a constant to specify the degree of gradient accumulation in order to match the execution time of forward propagations with their sub-model loading time. We empirically determine $C$ offline via profiling for a particular GPU.  



\noindent \textbf{Model partitioning algorithm.} Based on the augmented computation graph, we design an exhaustive layer-wise search algorithm to find a feasible partitioning that satisfies all constraints. Algorithm~\ref{alg:partition} shows the pseudo code for a feasible partitioning. The input to the algorithm is the computation graph $G$ of a GPT-3 model that is sorted according to the nodes' execution order on GPU. The output is a candidate partitioning stored in $partitions$.

The algorithm recursively evaluates the last edge of a computation graph (line 9) to see if a partitioning that generates the largest sub-model given the computation graph satisfies the constraints. The sub-model evaluated for computation is indexed by $c_s$ and $c_e$ while the sub-model tested for its loading time is indexed by $l_s$ and $l_e$. The search is performed with a $step\_size$ of 1. 

Graph partitioning is an NP-hard problem and an exhaustive search could be prohibitively expensive when the graph contains hundreds of thousands of nodes. Knowledge on the structure of GPT-3 helps accelerate the search and can be input to the algorithm as an additional constraint. GPT-3 contains identical transformer blocks and this domain knowledge greatly reduces the search space. Among the multiple feasible partitions returned by the algorithm, \atom selects the one that minimizes the total tensor sizes between sub-models. The time complexity of the given algorithm is exponential in the number of nodes in the input graph $G$, and also depends on $step\_size$ of the nested for-loops in the Main function. 

\noindent \textbf{Model compilation and code generation.} 
\atom does not require any changes to user code and automatically generates Python code for the partitioned model. The compiler runtime takes as input the augmented computation graph, the partitions returned by the search algorithm, and the original model (\texttt{torch.nn.Module}). The output of the code generation engine is Python source code of the partitioned with each sub-model being a separate class. The same Python source code is distributed to all \atom worker nodes in a decentralized network. The process of code refactoring can be automated.



\subsection{Model Distribution and Fault Tolerance}


Node failures and transient network disruptions are common in a large-scale decentralized network. Unlike model and parallelism that suffer from single point of failure or do not support dynamic node join/leave, \atom allows volunteer nodes to independently train a copy of the complete model and relies on an periodic allreduce communication to synchronize copies of the model, in a way similar to data parallelism. \atom employs a distributed hash table (DHT) to monitor the status of volunteer nodes. Each node periodically publish its status to the DHT via a heartbeat message. Among the information included the heartbeat, nodes report the number of mini-batches they have processed so far. \atom summarizes the total number of mini-batches processed by all nodes and triggers an allreduce communication if a global batch size has been reached. In cases when node failure or departure occurs, training can still proceed as long as other nodes are still able to process mini-batches and compensate the loss due to the leaving node. 


The shared GPU clusters contributed by volunteers can handle multiple training jobs simultaneously given enough host memory. Management of different training jobs submitted by volunteers require individual progress of their training jobs. In a shared environment with different training jobs, each job will be scheduled for a timeslot to execute. Participants publish the generated model code and store the model to the distributed hash table so the model can be retrieved without any central storage even volunteer hardware continuously join, leave, and fail. Also, the compiled model can be reused for the same hardware configuration. Therefore, Atom can coordinate multi-tasking globally by means of service orchestration which is a future direction of this paper.

\begin{figure*}[ptb!]
\centering 
\begin{subfigure}{0.25\textwidth}
  \includegraphics[width=\linewidth]{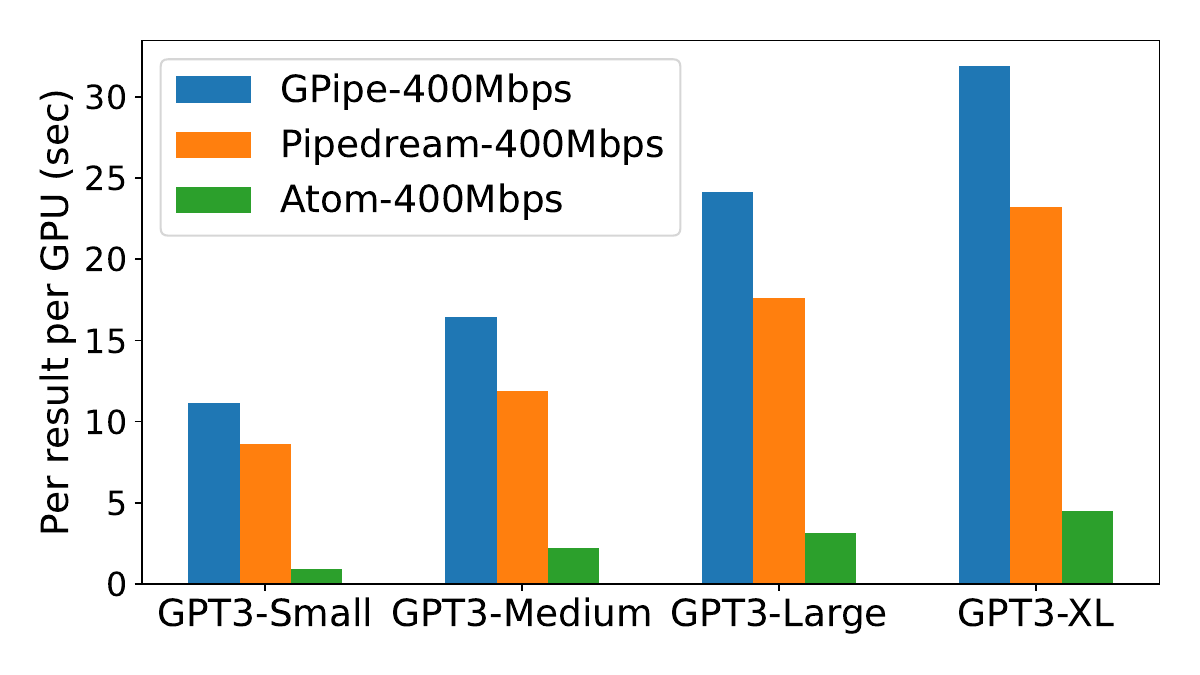}
  \caption{1080 Ti 400Mbps}
  \label{figs:gpt3_step_times_1080ti_400mbps}
\end{subfigure}\hfil
\begin{subfigure}{0.25\textwidth}
  \includegraphics[width=\linewidth]{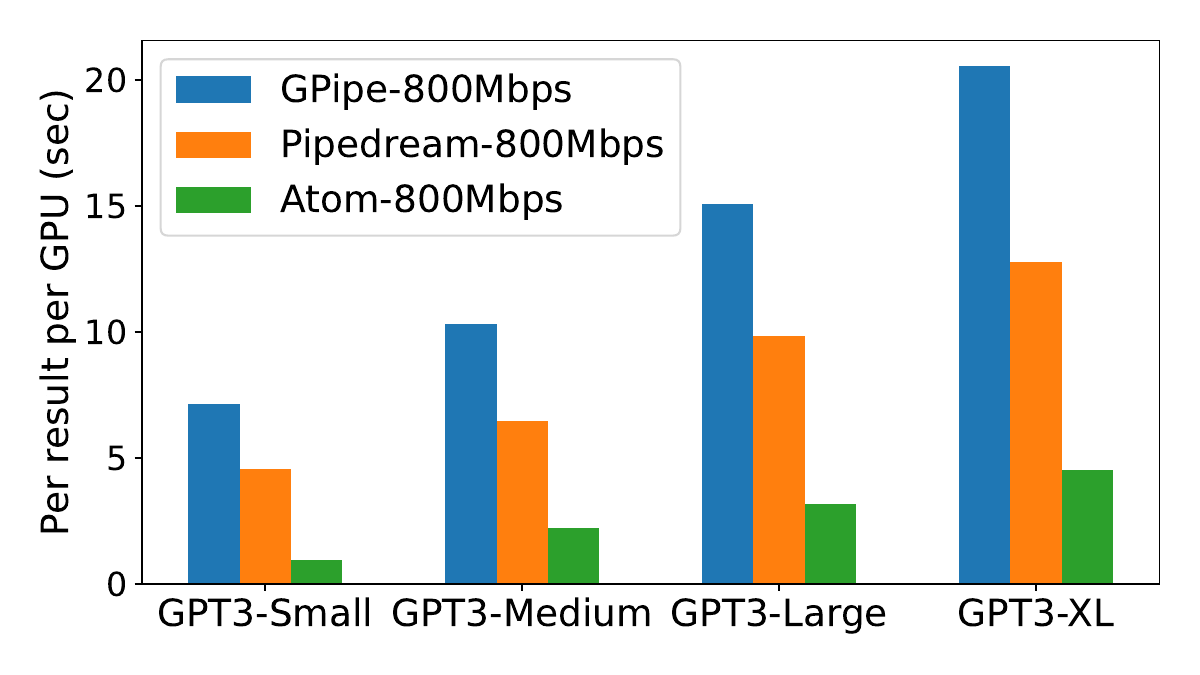}
  \caption{1080 Ti 800Mbps}
  \label{figs:gpt3_step_times_1080ti_800mbps}
\end{subfigure}\hfil 
\begin{subfigure}{0.25\textwidth}
  \includegraphics[width=\linewidth]{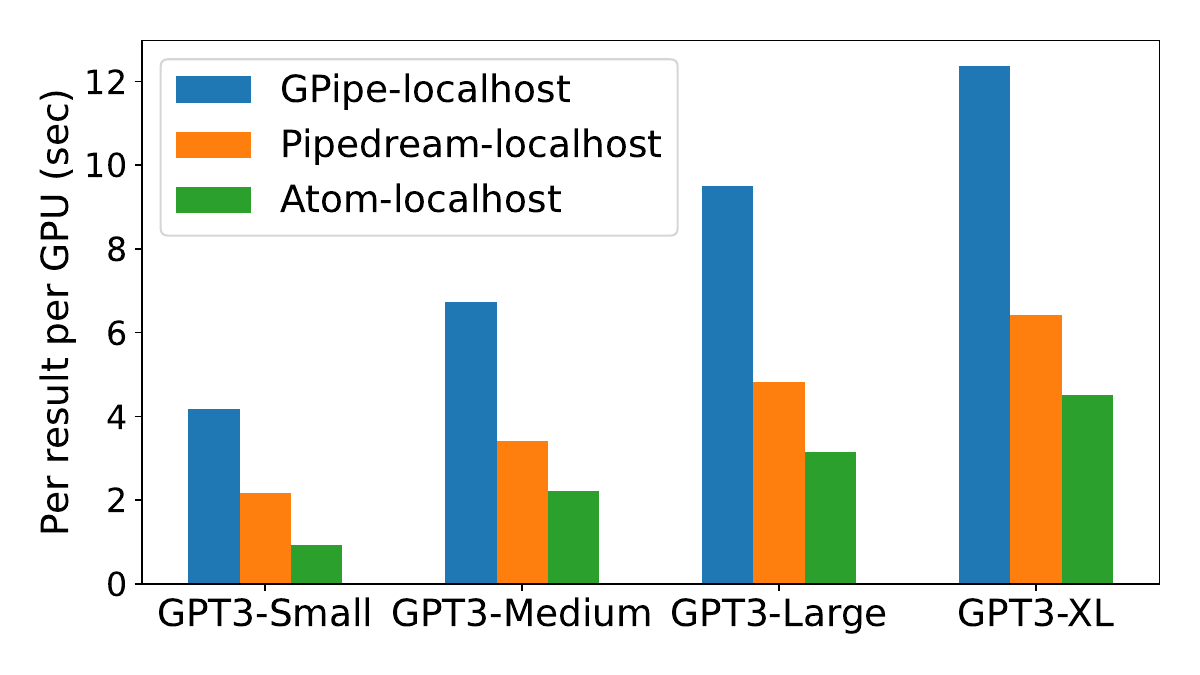}
  \caption{1080 Ti Localhost}
  \label{figs:gpt3_step_times_1080ti_localhost}
\end{subfigure}
\medskip 
\begin{subfigure}{0.25\textwidth}
  \includegraphics[width=\linewidth]{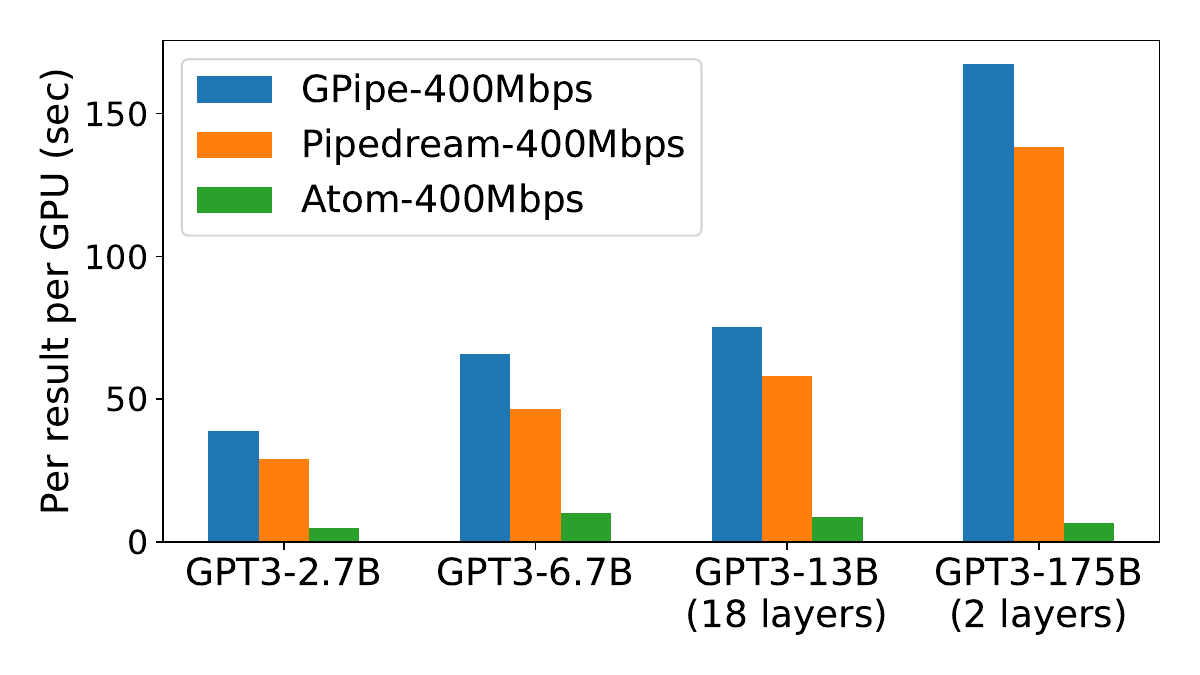} 
  \caption{V100 400Mbps}
  \label{figs:gpt3_step_times_v100_400mbps}
\end{subfigure}\hfil 
\begin{subfigure}{0.25\textwidth}
  \includegraphics[width=\linewidth]{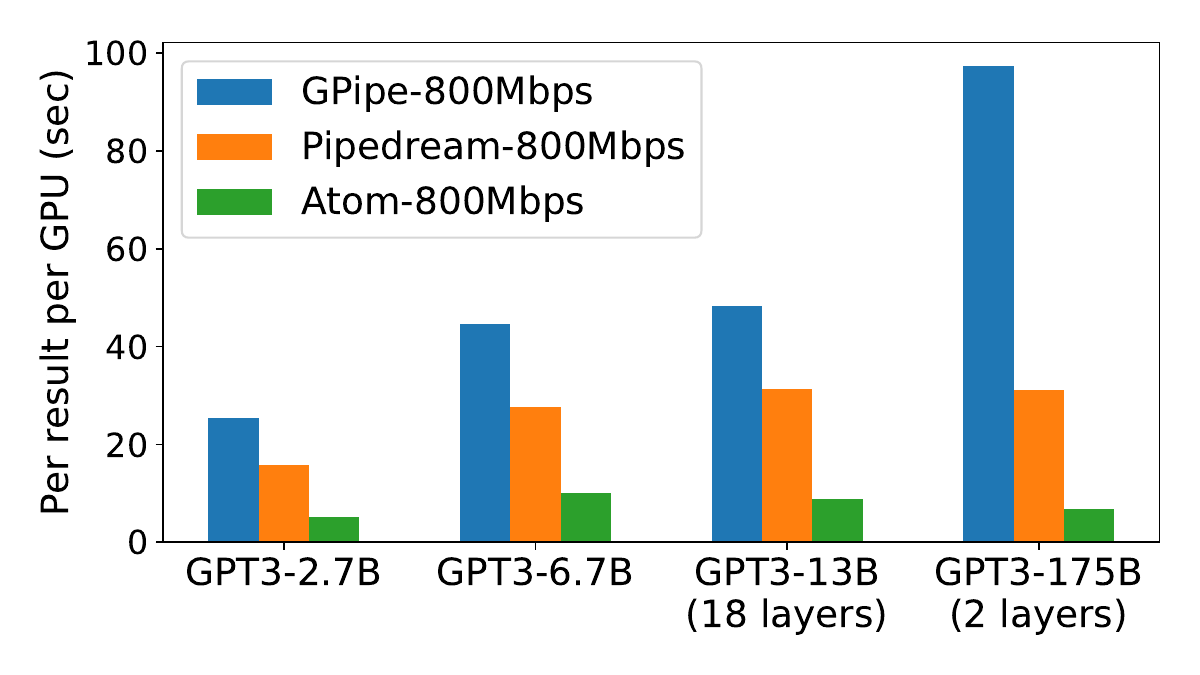}
  \caption{V100 800Mbps}
  \label{figs:gpt3_step_times_v100_800mbps}
\end{subfigure}\hfil  
\begin{subfigure}{0.25\textwidth}
  \includegraphics[width=\linewidth]{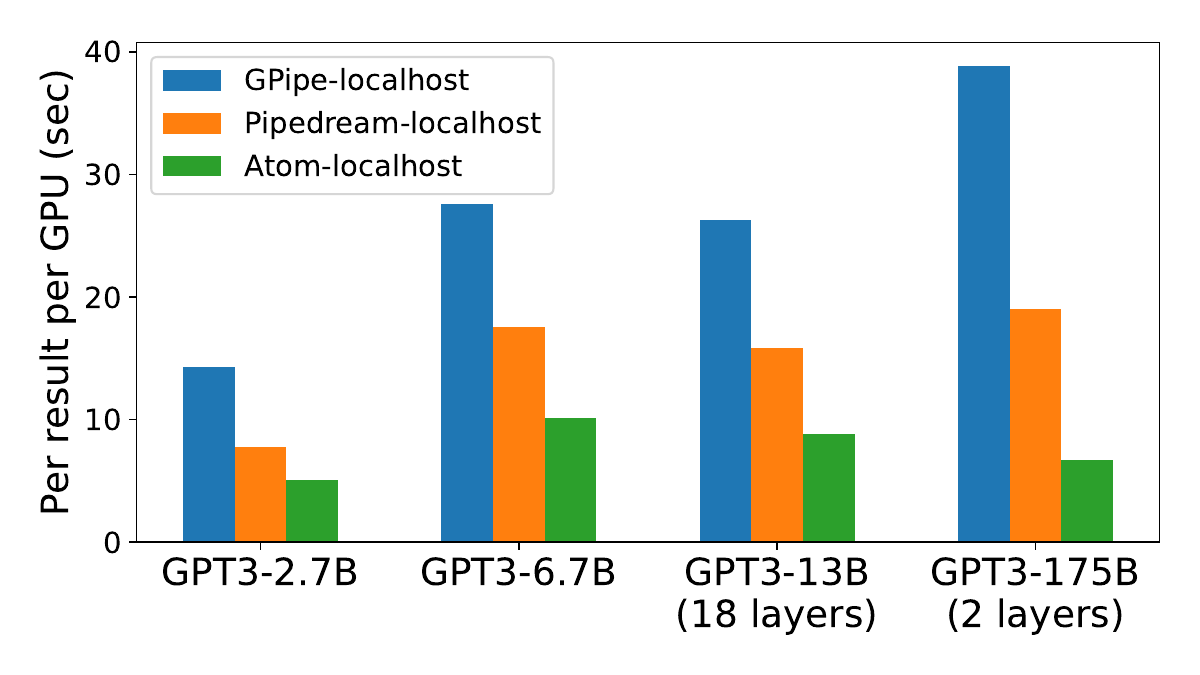}
  \caption{V100 Localhost}
  \label{figs:gpt3_step_times_v100_localhost}
\end{subfigure}
\caption{Comparison of per result (mini-batch) GPU time (w/o allreduce and optimizer step).}
\label{figs:step-time} 
\end{figure*}

\section{Implementation}

\atom is implemented based on the PyTorch framework with $\backsim$4K lines of code in Python. It uses the DHT implementation in Hivemind~\cite{hivemind} to coordinate volunteer nodes in a decentralized network and the code generation engine in FTPipe~\cite{eliad2021fine} to generate code for sub-models. The system consists of model tracing, profiling, model partitioning, and compilation. In profiling, a model is analyzed with a sample input and operators and layers are traced to construct a computation graph in the form of an internal representation. 


In the partitioning phase, we apply Algorithm~\ref{alg:partition} to assign partition id for each node based on the profiling statistics in the computation graph. At code generation phase, the compiler populates the initial function to define operators and layers of a sub-model, and define the execution flow in the \texttt{torch.nn.Module.forward} function base on the computation graph. The input shape is derived from the cutting edge between sub-models. During model execution, \atom maintains two CUDA streams. One is responsible for executing the current sub-model and the other one asynchronously prefetches the next sub-model immediately after the current sub-models starts. The former stream also swaps out a completed sub-model to host memory. Note that \atom does not swap out the last sub-model at the end of the forward pass or the first sub-model at the end of the backward pass to exploit locality between the forward and backward propagation.


\section{Evaluation}
\label{sec:evaluation}
In this section, we seek to answer the following questions:
\begin{itemize}
\item How well do model swapping perform on a single GPU in training LLMs compared with baseline pipeline parallelism strategies across servers in a low-bandwidth network environment?  (\S~\ref{eval:throughput})
\item How well does \atom scale with an increasing number of volunteer nodes?  (\S~\ref{eval:scalability-and-elasticity})
\item Does decentralized training in \atom affect the convergence of the trained model in the presence of node join and leave (fault tolerance) ? (\S~\ref{eval:convergence})
\end{itemize}

\subsection{Evaluation Methodology}

\noindent \textbf{Experimental setup.}
We use three types of GPU servers coded {\em high}, {\em medium}, and {\em low}. High is configured with four NVIDIA Tesla V100 GPUs, 72 cores (dual Intel Xeon Gold 6140 CPU@2.30GHz), and 385 GB host memory. Medium is configured with four GeForce GTX 1080 Ti GPUs, 72 cores (dual Intel Xeon CPU E5-2695 v4@2.10GHz), and 256 GB host memory. Low is configured with four GeForce GTX 1080 GPUs, 40 cores (dual Intel Xeon Silver 4114 CPU@2.20GHz), and 256 GB host memory. All servers have PCI-e 3.0x16. The CUDA Version is 11.3, and the PyTorch version is 1.11.0.

\noindent \textbf{Workloads.}
We mainly focus on GPT-3 models with various configurations~\cite{brown2020language}, including GPT-3 Small, Medium, Large, XL, 2.7B, 6.7B, 13B, 175B. Note that \atom requires that a model can fit in host memory. The full version GPT-3 175B model needs 2.8 TB memory and does not fit in our platforms. In general, GPT-3 175B can be trimmed down by removing a few identical transformer blocks to fit in our 385 GB host memory. For fair comparison with pipeline parallelism, in which the model size cannot exceed the aggregate GPU device memory, we evaluate a GPT-3 175B model with two transformer blocks and a model size of 68 GB.  

\noindent \textbf{Baseline.}
For each GPT3 model configuration, we compare \atom against Petals~\cite{borzunov2022petals} based on BigScience~\cite{BigScience} Project BLOOM~\cite{scao2022bloom} by using the schedule policy of GPipe~\cite{huang2019gpipe} and PipeDream~\cite{narayanan2019pipedream}. The baselines were also implemented in Hivemind~\cite{hivemind} similar as \atom to account for the effect of software architecture on training performance. Other state-of-the-art approaches are \emph{NOT} directly comparable since many assume the availability of high-speed interconnect. For example, ZeRO-Offload~\cite{ZeRO-Offload} has comparable performance of a single GPU swapping scheduling but it distributes optimization states, gradients, and parameters across GPUs, which can be inefficient over low-bandwidth networks. This is a significant contrast to our approach, which utilizes distributed decentralized training to avoid single-point failure. 

\noindent \textbf{Evaluation metrics.} We use per result GPU time, defined as the time to process one mini-batch per GPU, to measure training performance. Practically, we measure the number of mini-batches processed per GPU per unit time and take the reciprocal. We are also interested in the total time required to finish one training step, including the forward and backward propagation time, allreduce communication time, and the optimization time. We use the allreduce time to study \atomnospace's scalability. 

\subsection{Experimental Results}

\subsubsection{Training performance} \label{eval:throughput}

We first compare per result GPU time due to \atom and the two baselines, GPipe and PipeDream, with three network bandwidth settings: 400 Mbps, 800 Mbps, or localhost, the hypothetical upper bound on bandwidth. We throttled the bandwidth using Wonder Shaper~\cite{wondershaper}. The configurations are summarized in the \autoref{table:eval-config}. In this experiment, we disabled allreduce communication and the optimization step to focus only on the forward and backward passes. In all experiments, we used four GPUs on a single host without communications between hosts. This controlled environment helps compare the efficiency of the training pipeline in the baselines with that in \atomnospace. Models are optimally partitioned on four GPUs with minimal inter-sub-model tensor transmission for the baselines. \atom uses the automatically partitioned models by its search algorithm. We set the degree of gradient accumulation to 4 mini-batches. 

\begin{table}[h]
\setlength\tabcolsep{0pt} 
\caption{The configuration to compare training performance.}
\label{table:eval-config}
\begin{tabular*}{\columnwidth}{@{\extracolsep{\fill}} l*{3}{c}}
\toprule
\multicolumn{1}{c}{\textbf{Setting}} & \multicolumn{1}{c}{\textbf{Configuration}} \\
\midrule
\textbf{Schedule policy}        & Petals (GPipe, PipeDream), \atom \\
\textbf{Bandwidth}       & 400 Mbps, 800 Mbps, localhost \\
\textbf{Model config}    & GPT-3-Small, Medium, Large, XL, 2.7B, 6.7B, \\
                         & 13B (18 layers), 175B (2 transformer blocks) \\
\textbf{GPUs}            & 4$\times$1080 (8 GB), 4$\times$1080Ti (11 GB), \\
                         & 4$\times$V100 (32 GB) \\ 
\bottomrule
\end{tabular*}
\end{table}

\begin{figure}[ptb!]
\centering
\includegraphics[width=0.9\columnwidth]{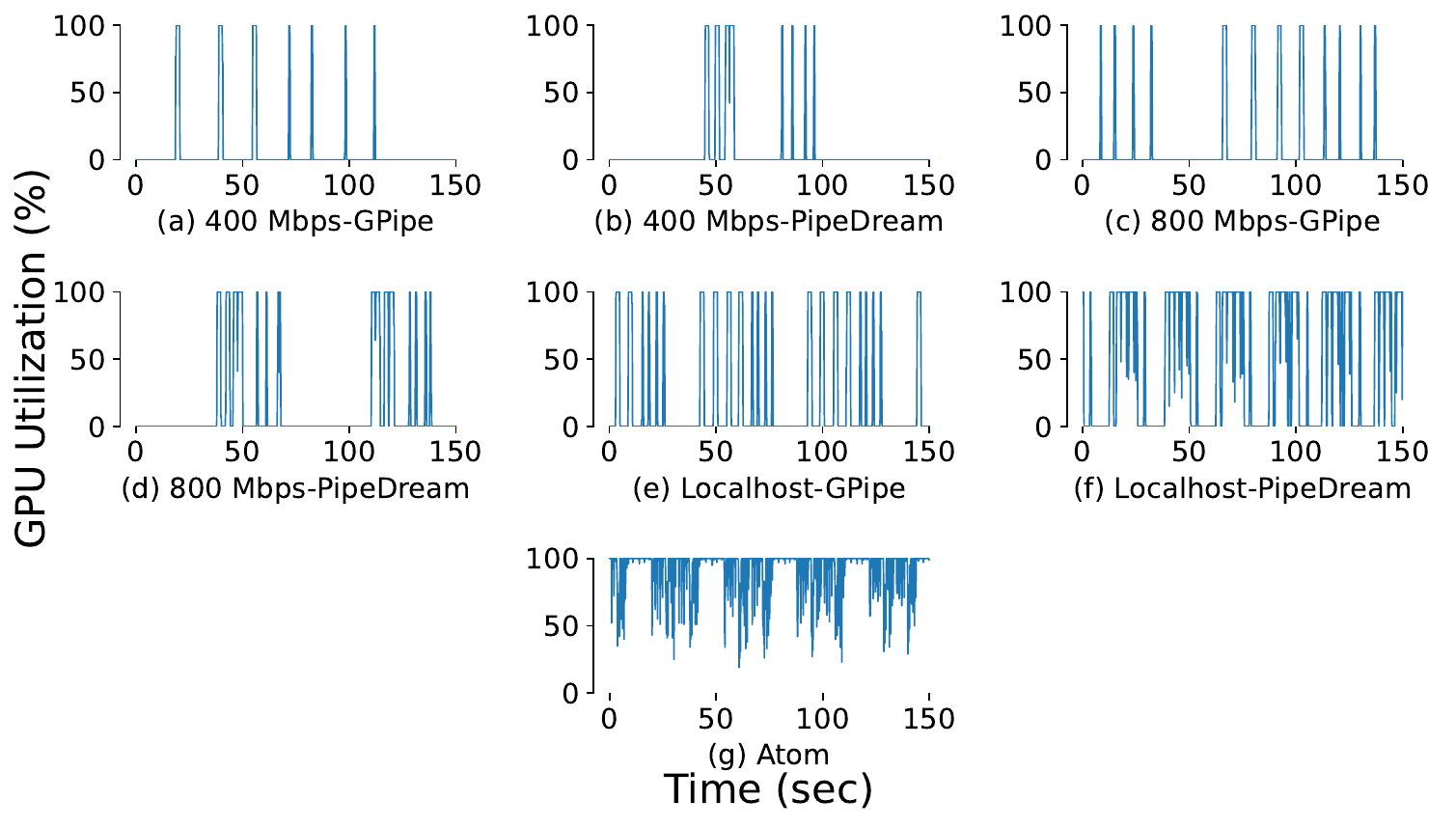}
\caption{GPU utilization over different network bandwidth of GPT-3 175B (2 transformer blocks) by \textit{nvidia-smi}.}
\label{figs:gpu_util}
\end{figure}


\begin{figure*}[ptb!]
\centering 
\begin{subfigure}{0.3\textwidth}
  \includegraphics[width=\linewidth]{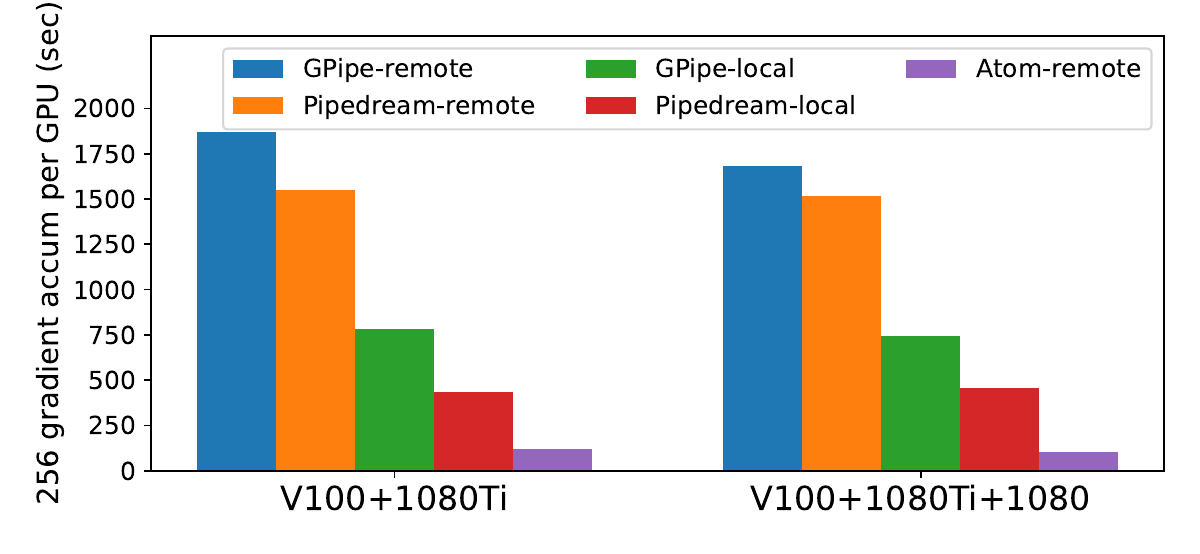}
  \caption{400 Mbps.}
  \label{figs:400mbps_allreduce_throughput_per_gpu}
\end{subfigure}\hfil
\begin{subfigure}{0.3\textwidth}
  \includegraphics[width=\linewidth]{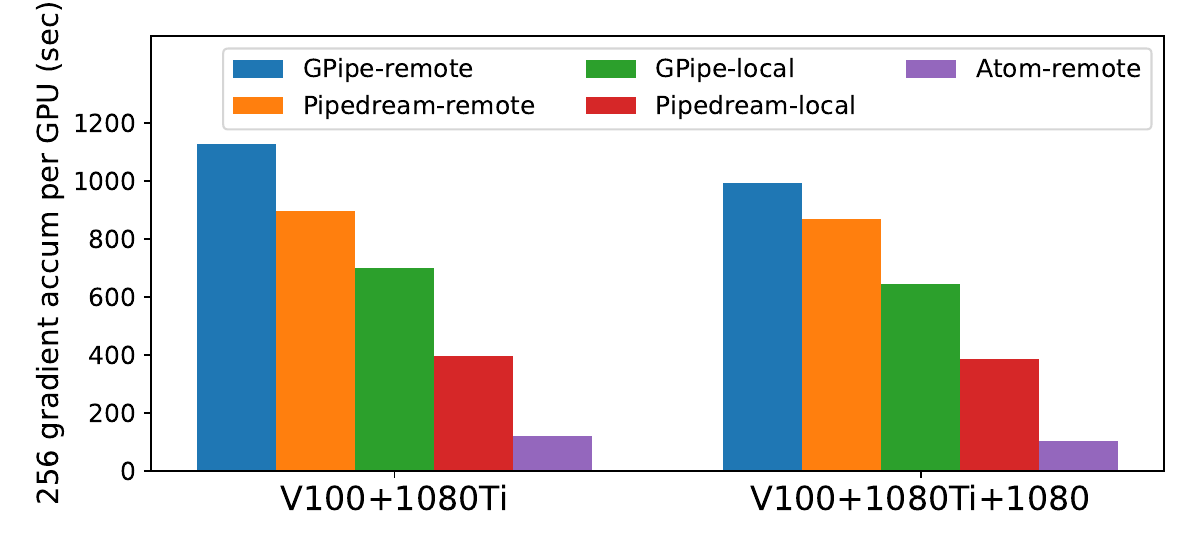}
  \caption{800 Mbps.}
  \label{figs:800mbps_allreduce_throughput_per_gpu}
\end{subfigure}\hfil 
\begin{subfigure}{0.3\textwidth}
  \includegraphics[width=\linewidth]{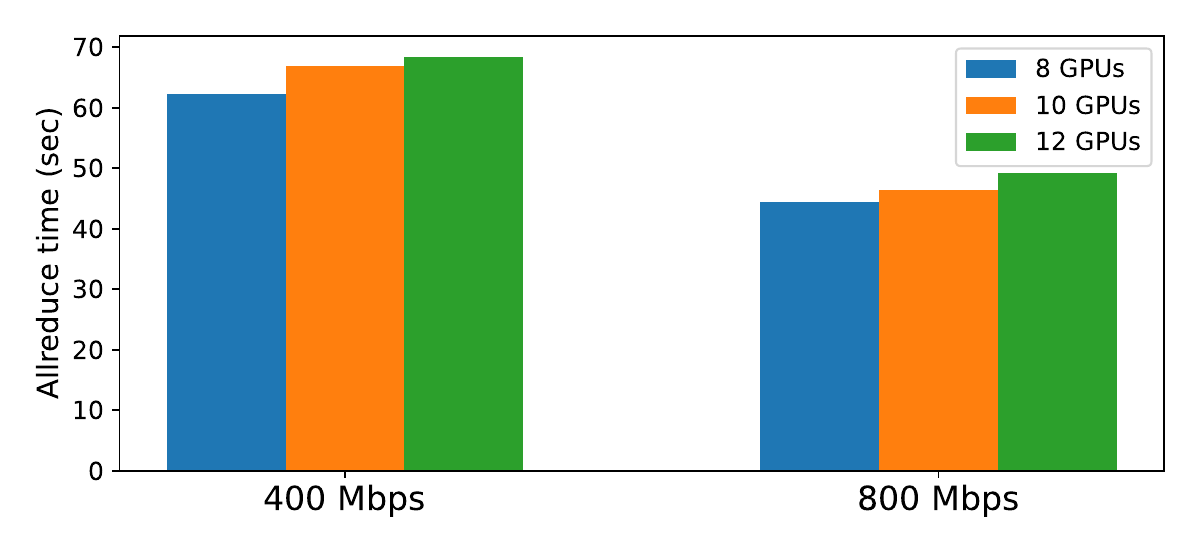}
  \caption{Change of allreduce time in \atom.}
  \label{figs:scaling}
\end{subfigure}
\caption{Comparison of total training time among GPipe, PipeDream, and \atom under different bandwidth.}
\label{figs:scalability} 
\end{figure*}

\autoref{figs:step-time} shows the averaged step time (both forward and backward) to process a mini-batch per GPU. Among the three approaches, \atom significantly and consistently outperformed GPipe and PipeDream in all experiments. The gap between \atom and the baselines widens as model size increases or the bandwidth decreases, suggesting that \atom is most suitable for training large-scale models with low-bandwidth networks. PipeDream also consistently outperformed GPipe due to its asynchronous training pipeline that is able to overlap the processing of multiple mini-batches. 

To further study the causes of differences in training performance, we plot GPU utilizations over the same period of time for \atom, GPipe, and PipeDream in \autoref{figs:gpu_util}. Both GPipe and PipeDream suffered from substantial idleness in GPU across all experiments. The idleness increases as network bandwidth drops. It suggests that even with multiple mini-batches being fed to the pipelines, computation cannot overlap with and hide inter-model communications, causing bubbles in the pipeline. In comparison, \atom is not affected by network speed and the automatically generated model partitioning helps streamline model execution and loading, leading to high GPU utilization. Both GPipe and PipeDream achieved the highest GPU utilization of 18.3\% and 46.3\%, respectively. with the localhost network. On contrary, \atom achieved a GPU utilization of 91.9\%, almost twice as much as PipeDream with an asynchronous pipeline. 

\subsubsection{Scalability} \label{eval:scalability-and-elasticity}
The scalability in distributed training is usually determined by the allreduce communication phase for model synchronization. For pipeline parallelism, there is a limit on how much a model can be partitioned to maintain a high computation-to-communication ratio. To scale to a large number of GPUs, a common practice is to have independent pipelines, each spanning multiple GPUs and servers. Similar to data parallelism, the pipelines are periodically synchronized via allreduce communication. To study \atomnospace's scalability, we focus on how is \atomnospace's performance compared with the baselines when using the same number of GPUs and the changes in its allreduce time as the number of GPU increases. 

Due to no access to a large GPU cluster, which motivated this work, we placed the three GPU servers, i.e., {\em high}, {\em medium}, and {\em low}, into three subnets connected by campus Internet to emulate a decentralized network. We configured the baselines with two settings and two scales. Suffix {\em local} indicates GPUs on the same host can communicate with unlimited bandwidth while {\em remote} indicates the bandwidth between any GPUs are throttled to emulate a wide-area network. We evaluated a 2-pipeline and 3-pipeline setting for the baselines. For example, the 2-pipeline experiment ran one pipeline on the V100 and 1080Ti server, respectively. For all experiment, \atom was configured to use the same type and number of GPUs as the baselines. All approaches were configured to perform an allreduce communication once a global batch size of 256 is reached. 

\autoref{figs:400mbps_allreduce_throughput_per_gpu} and \autoref{figs:800mbps_allreduce_throughput_per_gpu} show the comparison of the average training time that is needed to finish one global batch, i.e., 256 mini-batches. Note that in this experiment allreduce communication and optimizer updates were enabled. Again, \atom clearly outperformed the other two approaches by a large margin. It indicates that \atomnospace's decentralized architecture built on GPUs training on independent models is more scalable than the pipeline architecture that requires frequent inter-machine communication. \autoref{figs:scaling} shows the change in \atomnospace's allreduce time as the number of GPU increases. The results suggest that there is no dramatic change in the allreduce time as the system scaled. Because \atom consists of largely independent GPUs, its scalability is determined by the selection of an allreduce communication algorithm, which is orthogonal to \atomnospace's design. 

\begin{figure}[!ptb]
\setlength{\belowcaptionskip}{-0.2cm}           
\centering
\includegraphics[width=0.7 \columnwidth]{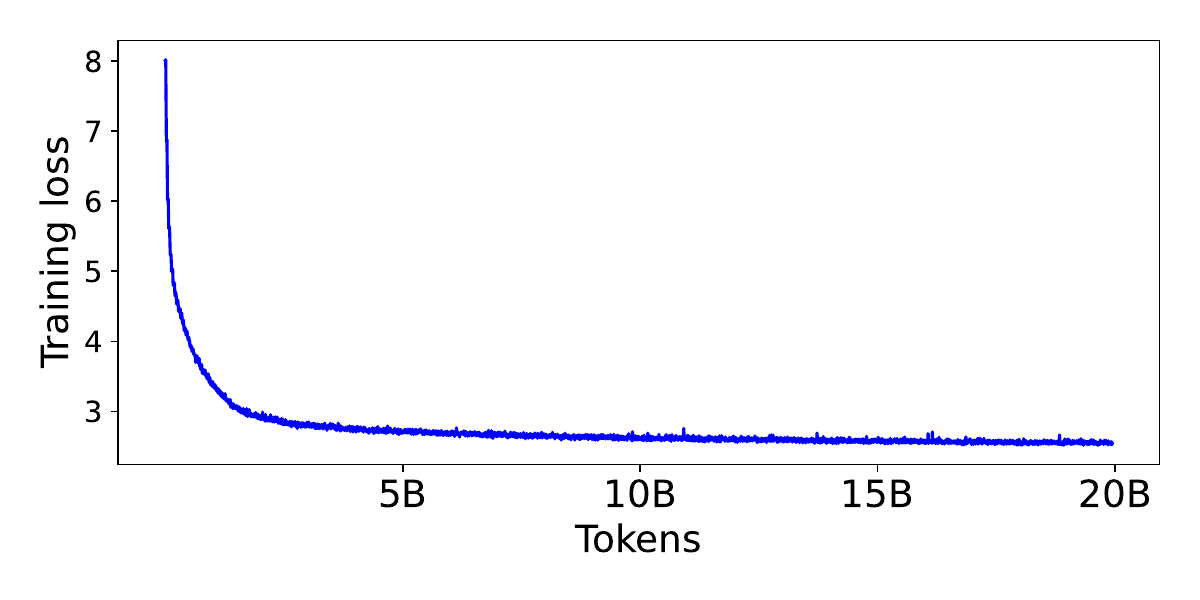}
\caption{Convergence of a GPT-3 Small model in \atomnospace.}
\label{figs:loss}
\end{figure}

\subsubsection{Effectiveness of training and fault tolerance} \label{eval:convergence}
\hfill\\ 
To improve training efficiency, \atom differs from the two baselines in training method. GPipe uses a synchronous pipeline that suffers from low GPU utilization but offers fast model convergence. PipeDream uses an asynchronous pipeline to overlap the training of multiple mini-batches. It improves GPU utilization but allows for updates on stale model parameters that results in slower convergence. In \atom, gradient accumulation is necessary for prolonging the forward pass to match its model loading time. However, gradient accumulation can negatively affect convergence. To verify \atomnospace's convergence, we trained a GPT-3 small model from scratch using the Wikipedia dataset~\cite{wikitext} until convergence and compare that with the baselines.  
The training used a learning rate $1 \times 10^{-4}$, 300K total training steps with a linear warmup of 3K steps. We used the CPU AdamW~\cite{loshchilov2017decoupled} optimizer ($\beta1$ = 0.9, $\beta2$ = 0.999). We set the target global gradient accumulation size to 512 samples. The target group size to do the model averaging (ring-allreduce step) and optimization is $12$. The total number of GPUs is 12 from 3 servers: 4$\times$V100, 4$\times$1080Ti, and 4$\times$1080. \autoref{figs:loss} shows the training loss over 20 billion tokens. We observe that the training loss steadily decreased and eventually converged to an accuracy comparable to the published result. We conclude that \atomnospace's architecture design for training efficiency does not undermine training effectiveness. During the experiment, we also deliberately killed two to four GPUs to emulate dynamically node joining/leaving. The training did not experience a disruption and was able to complete, though training performance dropped.

\section{Conclusion}


We present \atom for asynchronous training of massive models in a decentralized environment. The key insight that motivated \atom design is that huge LLMs can be executed on a single GPU layer by layer via memory swapping. \atom addresses the overhead of memory swapping by deriving an optimal schedule of model swapping through detailed profiling of individual layers of an LLM. The overarching objective that guides \atom design is to avoid GPU idleness as much as possible. Through comprehensive experiments, we demonstrate that the loosely-coupled distributed training architecture is advantageous than the tightly-coupled pipeline parallelism. We acknowledge that individual servers still needs to equip sufficient host memory to accommodate LLMs. We envision that with the development of the compute express link (CXL) interconnect, host memory will be more accessible than accelerators and high-speed networks.



\end{document}